\documentclass[useAMS,usenatbib]{mn2e}

\usepackage[dvips]{graphicx}
\usepackage[]{txfonts}
\usepackage{subfigure}
\usepackage{url}
\def\simless{\mathbin{\lower 3pt\hbox
{$\rlap{\raise 5pt\hbox{$\char'074$}}\mathchar"7218$}}}   %< or of order
\def\simmore{\mathbin{\lower 3pt\hbox
{$\rlap{\raise 5pt\hbox{$\char'076$}}\mathchar"7218$}}}   %> or of order
\newcommand{\be}{\begin{equation}}
\newcommand{\ee}{\end{equation}}
\newcommand       \bea          {\begin{eqnarray}}
\newcommand       \eea          {\end{eqnarray}}

\def\simlt{\mathrel{\hbox{\rlap{\hbox{\lower4pt\hbox{$\sim$}}}\hbox{$<$}}}}
\def\simgt{\mathrel{\hbox{\rlap{\hbox{\lower4pt\hbox{$\sim$}}}\hbox{$>$}}}}
\def\lesssim{\mathrel{\hbox{\rlap{\hbox{\lower4pt\hbox{$\sim$}}}\hbox{$<$}}}}

\def\gtrsim{\mathrel{\hbox{\rlap{\hbox{\lower4pt\hbox{$\sim$}}}\hbox{$>$}}}}

\title[]{Nuclear-dominated accretion and subluminous supernovae from the merger of a white dwarf with a neutron star or black hole}

\author[B.~D. Metzger]{B.~D. Metzger$^{1,2}$\thanks{E-mail:
bmetzger@astro.princeton.edu}\\
$^{1}$Department of Astrophysical Sciences, Peyton Hall, Princeton University, Princeton, NJ 08544, USA \\ $^{2}$NASA Einstein Fellow}

\begin{document}
\date{Received / Accepted}
\pagerange{\pageref{firstpage}--\pageref{lastpage}} \pubyear{2011}

\maketitle

\label{firstpage}

\begin{abstract}

We construct one dimensional steady-state models of accretion disks produced by the tidal disruption of a white dwarf (WD) by a neutron star (NS) or stellar mass black hole (BH).  At radii $r \lesssim 10^{8.5}-10^{9}$ cm the midplane density and temperature are sufficiently high to burn the initial white dwarf material into increasingly heavier elements (e.g.~Mg, Si, S, Ca, Fe, and Ni) at sequentially smaller radii.  When the energy released by nuclear reactions is comparable to that released gravitationally, we term the disk a nuclear-dominated accretion flow (NuDAF).  At small radii $\lesssim 10^{7}$ cm iron photo-disintegrates into helium and then free nuclei, and in the very innermost disk cooling by neutrinos may be efficient.  At the high accretion rates of relevance $\sim 10^{-4}-0.1M_{\odot}$ s$^{-1}$, most of the disk is radiatively inefficient and prone to outflows powered by viscous dissipation and nuclear burning.  Outflow properties are calculated by requiring that material in the midplane be marginally bound (Bernoulli constant $\lesssim 0$), due (in part) to cooling by matter escaping the disk.  For reasonable assumptions regarding the properties of disk winds, we show that a significant fraction ($\gtrsim 50-80\%$) of the total WD mass is unbound.  The composition of the ejecta is predominantly O, C, Si, Mg, Ne, Fe, and S [He, C, Si, S, Ar, and Fe], in the case of C-O [pure He] WDs, respectively, along with a small quantity $\sim 10^{-3}-10^{-2}M_{\sun}$ of radioactive $^{56}$Ni and, potentially, a trace amount of hydrogen.  Depending on the pressure dependence of wind cooling, we find that the disk may be thermally unstable to nuclear burning, the likelihood of which increases for higher mass WDs.  We use our results to evaluate possible electromagnetic counterparts of WD-NS/BH mergers, including optical transients powered by the radioactive decay of $^{56}$Ni and radio transients powered by the interaction of the ejecta with the interstellar medium.  We address whether recently discovered subluminous Type I supernovae result from WD-NS/BH mergers.  Ultimately assessing the fate of these events requires global simulations of the disk evolution, which capture the complex interplay between nuclear burning, convection, and outflows. 
\end{abstract}


\section{Introduction} 
\label{sec:intro}

Sensitive wide-field optical surveys are revolutionizing our understanding of time-dependent astrophysical phenomena.  New types of stellar explosions, that were once too faint or rare to be detected, are now routinely discovered by efforts such as the Palomar Transient Factory (\citealt{Law+09}; \citealt{Rau+09}) and PanSTARRs \citep{Kaiser+02}.  As our census of the transient universe expands, it becomes increasingly important to evaluate the expected electromagnetic counterparts of known astrophysical events.          

Several Type I supernovae (SNe) have recently been discovered that are dimmer and/or more rapidly evolving than normal SNe Ia or Ib/c.  Included among these is a class of peculiar Ia SNe characterized by lower luminosities and lower ejecta velocities than normal Ia's (`SN 2002cx-like events'; \citealt{Li+03}; \citealt{Jha+06}), and which occur predominantly in star-forming galaxies \citep{Foley+09}.  SN 2008ha is an extreme example, with a peak brightness and rise time of only $M_{\rm V} \simeq -14$ and $\sim 10$ days, respectively (\citealt{Valenti+09}; \citealt{Foley+09,Foley+10}).  Another recently identified class of subluminous transients are SNe Ib with Ca-rich [but S, Si, and Fe-poor] spectra (e.g.~SN 2005e; \citealt{Perets+10}).  These events appear to result from ejecta that have undergone helium burning (\citealt{Perets+10,Perets+11}; \citealt{Waldman+10}) and, unlike 2002cx-like events, are associated with an older stellar population located in the outskirts of their host galaxies.  Even more rapidly-evolving SNe such as 2002bj \citep{Poznanski+10} and 2010X \citep{Kasliwal+10} may be related to this class.

Lacking hydrogen in their spectra, Type I SNe are generally thought to originate from compact progenitors such as white dwarfs (WDs) or massive stars without extended envelopes.  Because compact stars lose most of their initial thermal energy to adiabatic expansion during the explosion, their emission must be powered by continued energy input from the radioactive decay of isotopes such as $^{56}$Ni.  The rapid evolution and low luminosities of events like 2002cx and 2005E thus require both a significantly lower $^{56}$Ni yield, and total ejecta mass, than characterize normal SNe.  

It is economical to associate new classes of SNe with anticipated variations of well-studied models, such as the core collapse of a massive star or the thermonuclear explosion of a WD.  Events like 2002cx may, for instance, result from the pure deflagration of a Chadrasekhar or sub-Chandrasekhar mass WD (e.g.~\citealt{Branch+04}; \citealt{Phillips+07}), as opposed to the `delayed detonation' models that best describe normal Ia SNe \citep{Nomoto+84}.  They may alternatively result from weak core collapse explosions (\citealt{Valenti+09}; \citealt{Moriya+10}).  For Ca-rich Ib SNe like 2005E, a promising explanation is the detonation of a helium shell on the surface of a C/O WD (\citealt{Woosley&Weaver86}; \citealt{Iben&Tutukov91}; \citealt{Livne&Arnett95}; \citealt{Bildsten+07}; \citealt{Shen+10}; \citealt{Woosley&Kasen10}).  Though promising, none of these explanations is yet definitive (e.g.~\citealt{Woosley&Kasen10}).  The possibility thus remains that at least some of these events represent an entirely new type of stellar explosion.  

In this paper we examine the observable signatures of the merger of a WD with a binary neutron star (NS) or black hole (BH) companion.  Four Galactic WD-NS binaries are currently known that will merge due to the emission of gravitational radiation within a Hubble time (see $\S\ref{sec:unstable}$).  If mass transfer following Roche Lobe overflow is unstable, then the WD is tidally disrupted, producing a massive disk around the NS or BH.  Past work has focused on the possibility that accretion powers a relativistic jet and a Gamma-Ray Burst (\citealt{Fryer+99}; \citealt{King+07}).  Here we instead focus on the properties of the accretion flow on larger scales.  We address two aspects of the problem neglected in previous studies: (1) the presence of [non-relativistic] outflows from the disk; and (2) the effects of nucleosynthesis on the thermodynamics and composition of the disk and outflows.  Although multidimensional numerical simulations are ultimately necessary to evaluate the evolution and fate of these systems, the large range in spatial and temporal scales involved make such a study numerically challenging.  Here, as a first step, we instead construct a simplified model of the disk and its outflows that we believe captures some of the essential physics.  We use our results for the mass, velocity, and composition of the ejecta to quantify the associated electromagnetic counterparts of WD-NS/BH mergers for a wide range of systems. 

This paper is organized as follows.  In $\S\ref{sec:unstable}$ we discuss the conditions for unstable mass transfer in WD-NS and WD-BH binaries and motivate the initial conditions for our subsequent calculations.  In $\S\ref{sec:model}$ we describe one dimensional steady-state models of the accretion disk and its outflows.  In $\S\ref{sec:results}$ we present our results for the disk structure ($\S\ref{sec:diskresults}$) and outflow properties ($\S\ref{sec:outflowresults}$).  In $\S\ref{sec:stability}$ we assess the stability of our solutions and in $\S\ref{sec:convection}$ we discuss the effects of convection.  In $\S\ref{sec:observe}$ we use our results to evaluate possible electromagnetic counterparts of WD-NS/WD-BH mergers, including subluminous Type I SNe ($\S\ref{sec:SN}$) and radio transients ($\S\ref{sec:radio}$).  We conclude in $\S\ref{sec:conclusion}$.

\section{Unstable Mass Transfer and Disk Formation}
\label{sec:unstable}

A standard scenario for the formation of tight WD-NS/BH binaries invokes common envelope evolution of an initially wide binary, consisting of a neutron star or black hole and a intermediate mass $\lesssim 8-10M_{\sun}$ main sequence companion (e.g.~\citealt{vandenHeuvel&Bonsdema84}).  For a limited range of orbital periods, unstable Roche lobe overflow (RLOF) begins only after the secondary has left the main sequence.  Depending on the evolutionary state of the stellar core when this occurs, the end result is a NS or BH in orbit with a WD with either (from lowest to highest WD mass) a pure He, He-C-O (`hybrid'), C-O, or O-Ne composition.  Close WD-NS/BH binaries can alternatively form directly by collisions in dense stellar regions, such as the centers of galaxies or globular clusters \citep{Sigurdsson&Rees97}.   

After the WD-NS/BH binary is brought close together, the system continues to lose angular momentum on a longer timescale to gravitational wave emission.  Angular momentum losses may be enhanced by the eccentricity induced by Kozai oscillations if a tertiary companion is present \citep{Thompson10}.  Of the $\gtrsim 20$ WD-NS binaries identified in our Galaxy \citep{Lorimer05}, only four are sufficiently compact that they will merge in $\lesssim 10^{10}$ yr (PSR J0751+1807 - \citealt{Kaspi+00}; J1757-5322 - \citealt{Edwards&Bailes01}; J1141-6545 - \citealt{Lundgren+95}; J1738-0333 - \citealt{Freire&Wex10}; see \citealt{OShaughnessy&Kim10} for a recent compilation).  The WD masses in these systems are $M_{\rm WD} = 0.18, 0.67, 0.99$, and 0.2$M_{\sun}$, respectively (e.g.~\citealt{Bailes+03}).  From the first three of these systems, \citet{Kim+04} estimate that the WD-NS merger rate in the Milky Way is $10^{-6}-10^{-5}$ yr$^{-1}$, although correcting for pulsar beaming increases this rate by a factor of several.  Population synthesis models predict somewhat higher rates $\sim 10^{-5}-10^{-3}$ yr$^{-1}$, but with larger uncertainty (\citealt{PortegiesZwart&Yungelson99}; \citealt{Tauris&Sennels00}; \citealt{Davies+02}).  No confirmed WD-BH systems are currently known.

Once the orbital period decreases to $\lesssim $ 1 minute, the WD undergoes RLOF onto the NS/BH companion.  Approximating the WD as a $\Gamma = 5/3$ polytrope, its radius is
\be
R_{\rm WD} \simeq 10^{4}\left(\frac{M_{\rm WD}}{0.7M_{\odot}}\right)^{-1/3}\left[1-\left(\frac{M_{\rm WD}}{M_{\rm ch}}\right)^{4/3}\right]^{1/2}\left(\frac{\mu_{e}}{2}\right)^{-5/3}{\,\rm km},
\label{eq:RWD}
\ee
where $M_{\rm ch} = 1.4M_{\odot}$ and $\mu_{e}$ is the mean molecular weight per electron \citep{Nauenberg72}.  The orbital separation at Roche contact is approximately \citep{Eggleton83}
\be
R_{\rm RLOF} = R_{\rm WD}\frac{0.6q^{2/3} + {\rm ln}(1+q^{1/3})}{0.49q^{2/3}},
\label{eq:rRLOF}
\ee
where $q = M_{\rm WD}/M$ and $M$ is the mass of the primary BH or NS.

If orbital angular momentum is conserved, then mass transfer is unstable for $q \gtrsim 0.4-0.55$ (e.g.~\citealt{Verbunt&Rappaport88}; see \citealt{Paschalidis+09}, their Fig.~11).  If strictly applicable, this criterion would limit tidal disruption to NS binaries with relatively massive $\gtrsim 0.5M_{\sun}$ C-O/O-Ne WDs, and might preclude disruption in WD-BH binaries altogether.  Stable systems increase their orbital separation after mass transfer begins.  The result is a long-lived accreting system, which may be observed as an ultracompact X-ray binary (e.g.~\citealt{Verbunt&vandenHeuvel95}).

Conservative mass transfer is, however, unlikely.  Tidal coupling during inspiral can transfer orbital angular momentum into spin.  Furthermore, the accretion rate onto the NS/BH just after RLOF is highly super-Eddington, in which case a common envelope may engulf the system and outflows are likely (e.g.~\citealt{Ohsuga+05}).  Depending on the amount of angular momentum lost to winds or tides, unstable mass transfer may occur in systems with even lower mass ratios $q \lesssim 0.3$ (e.g.~\citealt{Yungelson+02}).  In this case, tidal disruption may occur also for NS binaries with He or `hybrid' He-C-O WD companions, or even in the case of C-O/O-Ne WDs with low mass $\sim 3M_{\sun}$ BHs \citep{Fryer+99}.  Future numerical simulations are required to address the challenging issue of stability in degenerate binary mergers (e.g.~\citealt{Guerrero+04}; \citealt{DSouza+06}).

If mass transfer is unstable, then the WD is tidally disrupted in just a few orbits.  The WD material then circularizes and produces a disk around the central NS/BH with a total mass $M_{\rm WD}$ and characteristic radius $R_{\rm d} \sim R_{\rm RLOF}(1+q)^{-3}$.  This disk accretes onto the NS/BH on the viscous timescale
\begin{eqnarray}
&t_{\rm visc}& \simeq \alpha^{-1}\left(\frac{R_{\rm d}^{3}}{GM}\right)^{1/2}\left(\frac{H}{R_{\rm d}}\right)^{-2}\nonumber \\
&\sim& 140{\rm\,s} \left(\frac{\alpha}{0.1}\right)^{-1}\left(\frac{M}{1.4M_{\sun}}\right)^{-1/2}\left(\frac{R_{\rm d}}{10^{9}{\,\rm cm}}\right)^{3/2}\left(\frac{H/R_{\rm d}}{0.4}\right)^{-2}
\label{eq:tacc}
\end{eqnarray}
and at a characteristic rate
\begin{eqnarray}
&& \dot{M}(R_{\rm d})  \sim \frac{M_{\rm WD}}{t_{\rm visc}} \nonumber \\
&&\sim 10^{-2}M_{\sun}{\,\rm s^{-1}}\left(\frac{\alpha}{0.1}\right)\left(\frac{M}{1.4M_{\sun}}\right)^{1/2}\left(\frac{M_{\rm WD}}{0.7M_{\sun}}\right)\left(\frac{R_{\rm d}}{10^{9}{\,\rm cm}}\right)^{-3/2}\left(\frac{H/R_{\rm d}}{0.4}\right)^{2}\nonumber \\
\label{eq:mdot_out}
\end{eqnarray}
where $M$ is the NS/BH mass, $\alpha$ parametrizes the disk viscosity ($\S\ref{sec:model}$), and $H$ is the scale-height of the disk, normalized to a characteristic (\S\ref{sec:results}).  Turbulence in the disk is driven by the magneto-rotational instability, which transports angular momentum at a rate corresponding to $\alpha \sim 0.01-0.1$ (e.g.~\citealt{Davis+10}).  Because the mass of the disk is significant compared to that of the central NS or BH, the disk may also be gravitationally unstable, which likely results in even more efficient angular momentum transport (e.g.~\citealt{Laughlin&Bodenheimer94}).  In $\S\ref{sec:convection}$ we describe how the disk lifetime is altered if convection plays an important role in transporting angular momentum.  

Applying equations (\ref{eq:rRLOF}) and (\ref{eq:mdot_out}) to the full range of unstable systems, we conclude that the majority of the WD mass accretes at a rate $\dot{M}(R_{\rm d}) \sim 10^{-4}-10^{-1}M_{\sun}$ s$^{-1}$, depending primarily on $M_{\rm WD}$ and $\alpha$.  More massive WDs accrete at a higher rate, because both $\dot{M} \propto R_{\rm d}^{-3/2} \propto R_{\rm RLOF}^{-3/2}(1+q)^{9/2}$ and possibly $\alpha$ (if the disk is gravitationally unstable) increase with $q$.  At such high accretion rates ($\sim 10^{11}-10^{14}$ times larger than the Eddington rate) the disk cannot cool through photon emission and is termed a radiatively-inefficient accretion flow (RIAF).  

%-----------------------------------------------------------------  
\begin{figure}
\resizebox{\hsize}{!}{\includegraphics[angle=0]{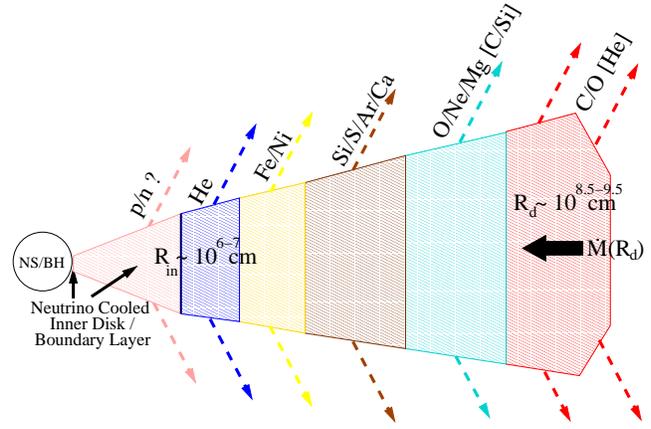}}
\caption[] {Schematic diagram of accretion and outflows following the merger of a WD-NS/BH binary.  At the outer edge of the disk (radius $R_{\rm d} \sim 10^{8.5}-10^{9.5}$ cm) the accretion rate $\dot{M}(R_{\rm d})$ is $\sim 10^{-4}-0.1M_{\sun}$ s$^{-1}$, depending on the properties of the merging binary and the viscosity $\alpha$ (eq.~[\ref{eq:mdot_out}]).  The outer composition is He, He-C-O, C-O, or O-Ne, depending on the properties of the disrupted WD.  As matter accretes to small radii, it experiences higher temperatures and burns to increasingly heavier elements.  Outflows are driven from the disk (directly or indirectly) by energy released from viscous dissipation and nuclear burning (Fig.~\ref{fig:CO}-\ref{fig:ONe}).  Elements synthesized near the midplane are transported to the surface, such that the local composition of the wind matches that of the midplane.  The outflow composition characterizing each range of radii are shown next to the outgoing arrows for the case of a C-O WD (the He WD case is shown in brackets).  The velocity of the wind is a factor $2\eta_{\rm w}^{1/2}$ times the local escape speed, such that the inner disk produces the highest velocity outflow.  On larger scales the ejecta forms a singular homologous outflow with a much lower velocity dispersion (see text).  At radii $r < R_{\rm in} \sim 10^{6}-10^{7}$ cm the temperature is sufficiently high that nuclei are photodissintegrated into $^{4}$He and, ultimately, free neutrons and protons.  Near the surface of the NS or BH, cooling by neutrinos may be important if the accretion rate is sufficiently high.
}
\label{fig:cartoon}
\end{figure}
%----------------------------------------------------------------

\section{Accretion Disk and Outflow Model}
\label{sec:model}

In this section we present a one dimensional steady-state model of the accretion disks produced by WD-NS/BH mergers.  Although the steady state approximation is clearly invalid just after the merger, it represents a reasonable description on timescales $\sim t_{\rm visc}$, during which most of the mass accretes at the rate estimated in equation (\ref{eq:mdot_out}).  We address the stability of our solutions in $\S\ref{sec:stability}$.

We adopt a height-integrated model, motivated by previous work on RIAFs in the context of `advection-dominated' models \citep{Narayan&Yi94} with outflows (\citealt{Narayan&Yi95}; \citealt{Blandford&Begelman99}).  Because RIAFs are geometrically thick, averaging over the vertical structure serves only as a crude approximation.  In standard RIAF models, viscous heating is balanced by advective cooling, due either to accretion or outflows (see \citealt{Narayan+98} for a review).  As we describe below, a key aspect of accretion following WD-NS/BH mergers is the relative importance of heating from nuclear reactions compared to viscous heating.  We term this novel accretion regime a Nuclear Dominated Accretion Flow (NuDAF).   

We begin by defining the (positive) steady state accretion rate
\be
\dot{M}(r) \equiv -2\pi r\Sigma v_{r},
\label{eq:mdotdisk}
\ee
where $\Sigma = 2H\rho$ is the surface density, $\rho$ is the midplane density, $H$ is the vertical scaleheight, and $v_{r} < 0$ is the radial velocity.  We allow for the presence of a wind with a mass loss rate $\dot{M}_{\rm w}$, which we characterize by the local quantity
\be
p(r) \equiv \frac{\partial{\rm ln\dot{M}_{\rm w}} }{\partial{\rm ln}r}
\label{eq:p}
\ee
We do not demand that $p(r)$ be constant with radius, but rather determine its functional form self-consistently from the solution, as described below.

We assume that the wind exerts no torque on the disk, such that it carries away only its own specific angular momentum, i.e.~$j_{\rm w} = j_{\rm d} = r^{2}\Omega$, where $\Omega$ is the angular velocity.  This is valid provided that the disk is not threaded by a large scale field of sufficient strength to establish an Alfven radius significantly above the disk surface.  We furthermore assume that no torque is applied by the inner boundary, such that (at radii much greater than the inner disk edge) the radial velocity obeys 
\be
v_{r} \simeq \frac{\nu}{\Omega}\frac{\partial \Omega}{\partial r} \approx -\frac{3}{2}\frac{\nu}{r} = -\frac{3}{2}\alpha\left(\frac{H}{r}\right)^{2}v_{\rm k},
\label{eq:vr}
\ee 
where we adopt a \citet{Shakura&Sunyaev73} kinematic viscosity $\nu = \alpha P/\rho\Omega$ with $\alpha \ll 1$; $H \approx a/\Omega < r$ is the vertical scale height; $a \equiv (P/\rho)^{1/2}$ defines the midplane sound speed; and in the last expression we assume that $\Omega$ equals the Keplerian rate $\Omega_{\rm k} = (GM/r^{3})^{3/2} = v_{\rm k}/r$ to leading order in $(H/r)^{2}$.  In adopting an anamalous viscosity we have implicitely assumed that angular momentum by MHD turbulence or gravitational instability {\it outwards} dominates any {\it inward} transport from convection; in $\S\ref{sec:convection}$ we discuss the validity of this point.  From equation (\ref{eq:mdotdisk}) we note that $\dot{M} = 3\pi\nu\Sigma$.

Mass continuity $\partial\dot{M}_{\rm d}/\partial r = \partial\dot{M}_{\rm w}/\partial r$ implies that
\be
\frac{\partial{\rm ln\rho}}{\partial{\rm ln}r} + 3\frac{\partial{\rm ln}a}{\partial{\rm ln}r} = p-3,
\label{eq:continuity}
\ee
while radial momentum conservation
\be
v_{r}\frac{\partial v_{r}}{\partial r} + \frac{1}{\rho}\frac{\partial P}{\partial r} + \frac{GM}{r^{2}} - r\Omega^{2} = 0
\ee
requires that
\be
\frac{\partial{\rm ln}\rho}{\partial{\rm ln}r} + 2\frac{\partial{\rm ln}a}{\partial{\rm ln}r} - \frac{r^{2}(\Omega^{2} - \Omega_{\rm k}^{2})}{a^{2}} = 0, 
\label{eq:radmomentum}
\ee
where we have neglected the first term $\propto v_{r}^{2} \propto (H/r)^{4}$.  

The entropy equation reads
\begin{eqnarray}
\dot{q}_{\rm adv} \equiv v_{r}T\frac{\partial s}{\partial r} = v_{r}\left(\frac{\partial\epsilon}{\partial r} + P\frac{\partial \rho^{-1}}{\partial r}\right) =  \frac{v_{r}a^{2}}{r}\left(\frac{2}{\gamma-1}\frac{\partial {\rm ln}a}{\partial {\rm ln}r} - \frac{\rm\partial ln\rho}{\partial {\rm ln}r}\right) = \dot{q} \nonumber \\ 
\label{eq:entropy}
\end{eqnarray}
where $T$ is the midplane temperature, $\gamma \in [4/3,5/3]$ is the adiabatic index, and $s$ and $\epsilon$ are the specific entropy and internal energy, respectively.  We have neglected the chemical potential term $\propto \mu_{i}\frac{\partial Y_{i}}{\partial r}$ because all species (including electrons) are non-degenerate at the radii where nuclear burning commences.  Equation (\ref{eq:entropy}) may be interpreted as the balance between ``cooling'' from advection through the disk $\dot{q}_{\rm adv}$ and others sources of heating/cooling $\dot{q} = \dot{q}_{\rm visc} + \dot{q}_{\rm nuc} + \dot{q}_{\rm w}$, including those resulting from nuclear reactions $\dot{q}_{\rm nuc}(\rho,T)$ (see eq.~[\ref{eq:qdotwind}] below), winds $\dot{q}_{\rm w}$ (see below), and viscous dissipation
\be
\dot{q}_{\rm visc} = \nu r^{2}\left(\frac{\partial \Omega}{\partial{\rm ln}r}\right)^{2} \simeq \frac{9}{4}\nu\Omega^{2} = \frac{9}{4}\nu\Omega_{\rm k}^{2}\left[1 + \left(\frac{H}{r}\right)^{2}\left(\frac{\partial{\rm ln}\rho}{\partial{\rm ln}r} + 2\frac{\partial{\rm ln}a}{\partial{\rm ln}r}\right)\right],
\label{eq:qdotvisc}
\ee
where we have used equation (\ref{eq:radmomentum}) in the last expression and neglect terms $\propto \partial(H/r)/\partial r$.  We again note that radiative cooling may be neglected because the timescale for photon diffusion from the disk midplane is much longer than the viscous time time (eq.~[\ref{eq:tacc}]).  In $\S\ref{sec:convection}$ we describe a scenario in which this might be violated, but this alternative picture of the disk evolution is radically different from the model developed in this section.    

An important characteristic of RIAFs without outflows ($p=0$) is that the Bernoulli parameter
\be
{\rm Be}_{\rm d} \equiv \frac{1}{2}v_{r}^{2} + \frac{1}{2}r^{2}\Omega^{2} + \epsilon + \frac{P}{\rho} -\frac{GM}{r}
\label{eq:Bed}
\ee
is generically positive \citep{Narayan&Yi94}.  This implies that, in principle, material in the midplane has sufficient energy to escape to infinity.  This fact has been used previously to argue that RIAFs are susceptible to powerful outflows (\citealt{Narayan&Yi95}; \citealt{Blandford&Begelman99}) that carry away a substantial portion of the accreting mass, viz.~$p \in [0,1]$.  If a fraction of the accreting matter does in fact escape, then the upper atmosphere of the disk must be {\it preferentially heated}.  Only in this manner can the wind achieve a positive Bernoulli parameter Be$_{\rm w} = v_{\rm w}^{2}/2 > 0$, while the remaining disk material is accordingly cooled and remains bound (Be$_{\rm d} \lesssim 0$), where $v_{\rm w}$ is the asymptotic velocity of the wind.  

Following previous work (e.g.~\citealt{Kohri+05}), we quantify the efficiency of wind heating by a parameter $\eta_{\rm w} \equiv v_{\rm w}^{2}/2v_{k}^{2}$, which equals the ratio of Be$_{\rm w}$ to the local gravitational binding energy.  This prescription is certainly not unique; for instance, $v_{\rm w}$ could instead scale with the disk sound speed $a$.  Although for our purposes here this distinction is unimportant since $v_{k}$ and $a$ have similar values, the pressure-dependence of wind cooling is different between these cases, which has an important effect on the thermal stability of the disk ($\S\ref{sec:stability}$).  

As we describe below, by demanding that the Bernoulli parameter in the midplane Be$_{\rm d} \leq 0$ and fixing $\eta_{\rm w}$, this uniquely specifies the wind mass outflow rate $p(r)$ (eq.~[\ref{eq:p}]).  Although we adopt a `two zone' model (disk + outflow), we do not specify the source of wind heating explicitly.\footnote{One possible source of coronal heating is the dissipation of MHD waves.  Waves may be excited by turbulence due to the MRI or vertical convection, the latter of which is likely due to the strong temperature dependence of the nuclear heating rate.}  In most of our calculations we adopt values for $\eta_{\rm w} \sim $ few, because a terminal speed of the order of the escape speed is a common feature of thermally-driven winds (e.g.~\citealt{Lamers&Cassinelli99}).

If the wind is heated, then the disk necessarily cools at the rate
\be
\dot{q}_{\rm w} = -\frac{(v_{\rm w}^{2}/2 - {\rm Be_{ d}})}{2\pi\Sigma r}\frac{\partial M_{\rm d}}{\partial{\rm ln}r} = -\frac{3}{2}p(\eta_{\rm w}-{\rm Be}_{\rm d}')\nu\Omega_{\rm k}^{2},
\label{eq:qdotwind}
\ee
where Be$_{\rm d}' \equiv {\rm Be}_{\rm d}/v_{k}^{2}$ is the `normalized' disk Bernoulli function.  

By combining equations (\ref{eq:entropy})$-$(\ref{eq:qdotwind}) we find that
%\begin{eqnarray}
%-\left(\frac{H}{r}\right)^{2}\left[a^{-2}\left.\frac{\partial\epsilon}{\partial {\rm ln}\rho}\right|_{a} + \frac{1}{2}\right]\frac{\partial{\rm ln}\rho}{\partial{\rm ln}r} - \left(\frac{H}{r}\right)^{2}\left[a^{-2} \left.\frac{\partial\epsilon}{\partial {\rm ln}a}\right|_{\rho} + 3\right]\frac{\partial{\rm ln}a}{\partial{\rm ln}r} \nonumber \\ = \frac{2}{3}\frac{\dot{q}_{\rm nuc}}{\nu\Omega_{k}^{2}} + \frac{3}{2} - p\eta_{\rm w}
%\end{eqnarray}
\begin{eqnarray}
\left(\frac{H}{r}\right)^{2}\left[\frac{3\gamma-1}{\gamma-1}\frac{\partial{\rm ln}a}{\partial{\rm ln}r} + \frac{1}{2}\frac{\partial{\rm ln}\rho}{\partial{\rm ln}r}\right] = -\frac{2}{3}\frac{\dot{q}_{\rm nuc}}{\nu\Omega_{k}^{2}} + p(\eta_{\rm w}-{\rm Be}_{\rm d}')-\frac{3}{2}
\label{eq:equation2}
\end{eqnarray}

Finally, nuclear reactions change the composition of the accreting material according to
\be
v_{r}\frac{\partial X_{A}}{\partial r} = \dot{X}_{A}|_{P} 
\label{eq:Xdot}
\ee
where $X_{A}$ is the mass fraction of isotope with mass number $A$, and $\dot{X}_{A}|_{P}$ is the nuclear reaction rate.  Equation (\ref{eq:Xdot}) shows that material burns for approximately the local accretion timescale $\sim r/v_{r}$ at any radius.  We assume that burning occurs at constant pressure because in general the burning timescale is long compared to the dynamical timescale over which vertical pressure balance is established.

We neglect the effects of convective/turbulent mixing in the disk, which may act to smooth radial abundance gradients; mixing could, in principle, be modeled by including an additional term $\propto \nu_{\rm mix}\nabla^{2}X_{A}$ to the right hand side of equation (\ref{eq:Xdot}), where $\nu_{\rm mix}$ is the diffusion coefficient.  Neglecting mixing is justified as a first approximation because compositional changes typically occur over a radial distance $\gtrsim H$.  Studies of the diffusion of passive contaminants in numerical simulations of the MRI furthermore demonstrate that $\nu_{\rm mix} < \nu$ (e.g.~\citealt{Carballido+05}), such that it appears unlikely that burned material will diffuse upstream.

The nuclear reaction rates $\dot{X}_{A}(\rho,T)$ and nuclear heating rate $\dot{q}_{\rm nuc}(\rho,T)$ in equations (\ref{eq:entropy}) and (\ref{eq:Xdot}) are calculated using a 19 isotope reaction network\footnote{See\,\url{http://www.cococubed.com/code_pages/burn_helium.shtml}} that includes $\alpha$-capture, heavy-nuclei, and ($\alpha$,p)(p,$\alpha$) reactions \citep{Timmes99}.  The temperature $T(r)$ and adiabatic index $\gamma(r)$ are calculated using a standard equation of state, which includes ideal gas, radiation, and degeneracy pressure.

We calculate solutions by integrating equations (\ref{eq:continuity}), (\ref{eq:equation2}), (\ref{eq:Xdot}) from the outer boundary $r = R_{\rm d}$ inwards, in order to obtain $a(r)$, $\rho(r)$ and $X_{A}(r)$.  The wind mass loss rate $p(r)$ (eq.~[\ref{eq:p}]) is determined by demanding that the Bernoulli function is regulated by outflows to a fixed value Be$_{\rm d}' = -0.2-0$ (marginally bound disk).  Boundary conditions are determined by fixing the outer accretion rate $\dot{M}_{\rm d}(R_{\rm d})$ (eq.~[\ref{eq:mdot_out}]) and by assuming that the inflowing material reaches a self-similar evolution with $a \propto v_{k} \propto r^{-1/2}$, such that $\frac{\partial{\rm lna}}{\partial{\rm lnr}}|_{R_{\rm d}} = -1/2$.\footnote{Regardless of the precise outer boundary condition adopted, the self-similar solution rapidly obtains (\citealt{Narayan&Yi94}).  Self-similarity is maintained until nuclear burning becomes important, which introduces an additional scale into the problem.}  We terminate our solutions at the radius $R_{\rm in} \lesssim 10^{7}$ cm at which heavy nuclei are photo-disintegrated into free nuclei.  Interior to this point, neutrinos from the reactions $e^{-}+p\rightarrow n + \nu_{e}$ and $e^{+} + n \rightarrow p + \bar{\nu}_{e}$ become an important source of cooling (e.g.~\citealt{Narayan+01}; \citealt{Chen&Beloborodov07}).  Depending on $\dot{M}$ and $\alpha$, the disk may become radiatively efficient and geometrically thin, such that strong outflows may be suppressed close to the surface of the central NS/BH.  Figure \ref{fig:cartoon} shows a schematic diagram of the structure of the disk and outflows.

The composition at the outer boundary $r = R_{\rm d}$ is that of the tidally disrupted WD.  For C-O WDs we adopt an initial composition $X_{12} = 0.5$ and $X_{16} = 0.5$ \citep{Salaris+97}.  We also perform calculations for O-Ne WDs ($X_{16} = 0.7$; $X_{20} = 0.3$; e.g.~\citealt{Gutierrez+05}), hybrid He-C-O WDs ($X_{4} = 0.2$; $X_{12} = 0.4$; $X_{16} = 0.4$; \citealt{Han+00}), and pure He WDs ($X_{4} = 1$), although the latter may occur less frequently in Nature due to the limited parameter-space available for unstable mass transfer ($\S\ref{sec:unstable}$).  Although we focus on WD-NS binaries, we also consider $3M_{\odot}$ BH companions.  Table \ref{table:models} enumerates the models calculated in the next section.
 
\section{Results}
\label{sec:results}
Figures \ref{fig:CO}-\ref{fig:ONe} and Tables \ref{table:models} and \ref{table:results} summarize the results of our calculations.  In this section we describe the disk structure ($\S\ref{sec:diskresults}$), outflow composition ($\S\ref{sec:outflowresults}$), and thermal stability of our solutions ($\S\ref{sec:stability}$).  In $\S\ref{sec:convection}$ we discuss the possible effects of convection on our results and in $\S\ref{sec:analytic}$ we present an analytic estimate of the importance of nuclear burning.
\subsection{Disk Structure}
\label{sec:diskresults}

\begin{figure}
\centering
\subfigure[Radial Profile of Disk Properties]{
\includegraphics[width=0.46\textwidth]{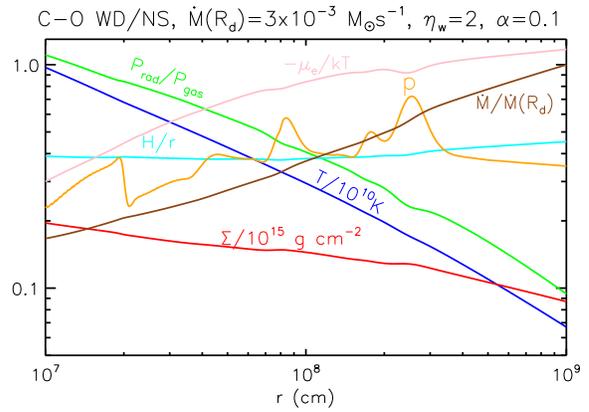}
}
\subfigure[Heating/Cooling Terms in the Entropy Equation]{
\includegraphics[width=0.46\textwidth]{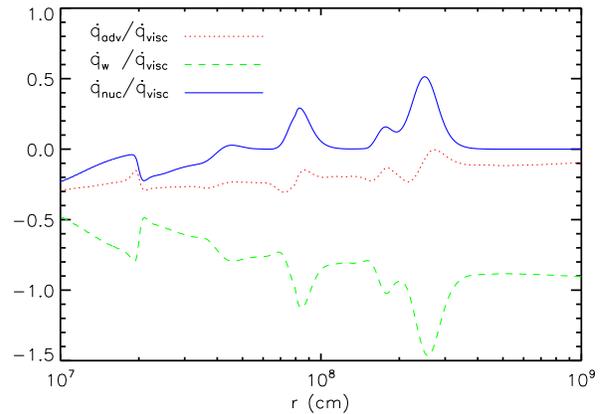}
}
\subfigure[Composition Profile]{
\includegraphics[width=0.46\textwidth]{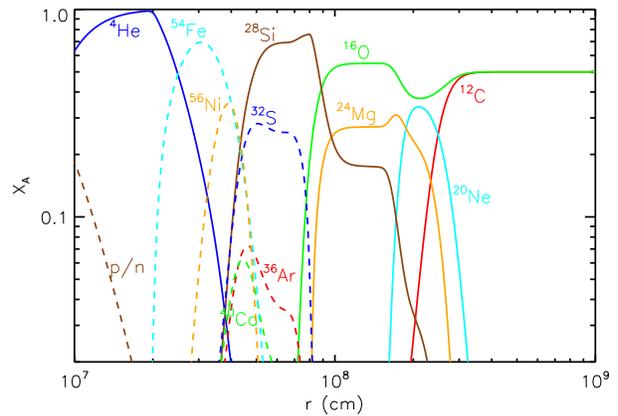}
}
\caption{Accretion following the merger of a 0.6$M_{\odot}$ WD with a 1.4$M_{\odot}$ NS.  We adopt fiducial values $\eta_{\rm w} = 2$ for the wind efficiency, Be$_{\rm d}^{'} = -0.1$ for the Bernoulli function of the disk (eq.~[\ref{eq:Bed}]), and $\alpha = 0.1$ for the disk viscosity.  The latter translates into an accretion rate $\dot{M}(R_{\rm d}) \simeq 3 \times 10^{-3}M_{\odot}$ s$^{-1}$ at the outer edge of the disk $R_{\rm d} = 10^{9}$ cm (eq.~[\ref{eq:mdot_out}]).  {\it \bf Top:} Disk temperature $T$ ({\it Blue}), surface density $\Sigma$ ({\it Red}), aspect ratio $H/r$ ({\it Turquoise}), ratio of radiation to gas pressure P$_{\rm rad}$/P$_{\rm gas}$ ({\it Green}), negative of the electron chemical potential $\mu_{e}$ divided by $kT$ ({\it Pink}), accretion rate $\dot{M}$ normalized to $\dot{M}(R_{\rm d})$ ({\it Brown}), and wind mass loss index $p$ ({\it Orange}; eq.~[\ref{eq:p}]), as a function of radius.  {\it \bf Middle:} Sources of heating and cooling (eq.~[\ref{eq:entropy}]), in ratio to the viscous heating rate $\dot{q}_{\rm visc}$: nuclear reactions $\dot{q}_{\rm nuc}$ ({\it Solid Blue}); advective cooling within the disk $\dot{q}_{\rm adv} \equiv Tv_{r}(\partial s/\partial r)$ ({\it Solid Dotted}); and wind cooling $\dot{q}_{\rm w}$ ({\it Dashed Green}; eq.~[\ref{eq:qdotwind}]).  {\it \bf Bottom:} Mass fraction $X_{A}$ of various isotopes in the disk midplane.
%, including $^{4}$He ({\it Solid Blue}), $^{12}$C ({\it Solid Red}), $^{16}$O ({\it Solid Green}%), $^{20}$Ne ({\it Solid Light Blue}), $^{24}$Mg ({\it Solid Orange}), $^{28}$Si ({\it Solid Bro%wn}), $^{32}$S ({\it Solid Blue}), $^{36}$Ar ({\it Dashed Red}), $^{40}$Ca ({\it Dashed Green}),% $^{52}$Fe ({\it Dashed Ligh Blue}), $^{56}$Ni ({\it Dashed Orange}), and free nucleons ({\it Dashed Brown}).
}
\label{fig:CO}
\end{figure}

\begin{figure}
\centering
\subfigure[Radial Profile of Disk Properties]{
\includegraphics[width=0.46\textwidth]{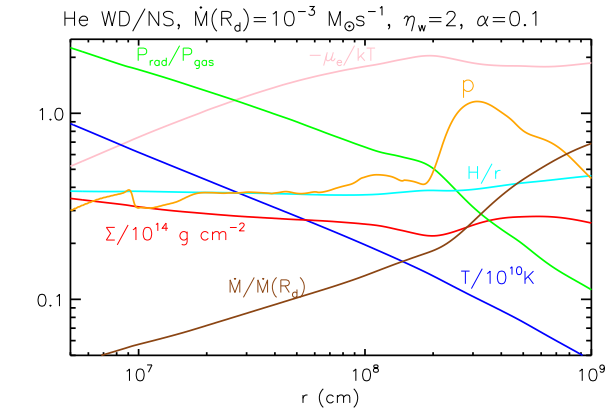}
}
\subfigure[Heating/Cooling Terms in the Entropy Equation]{
\includegraphics[width=0.46\textwidth]{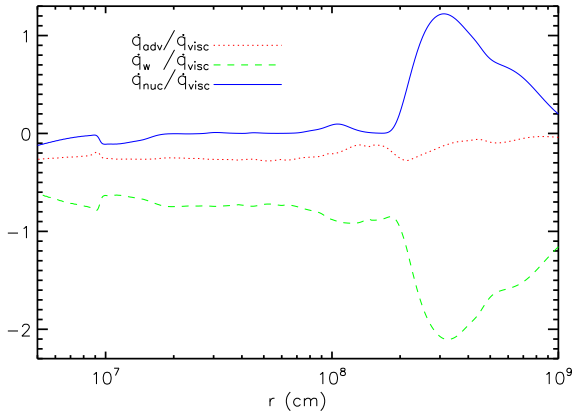}
}
\subfigure[Composition Profile]{
\includegraphics[width=0.46\textwidth]{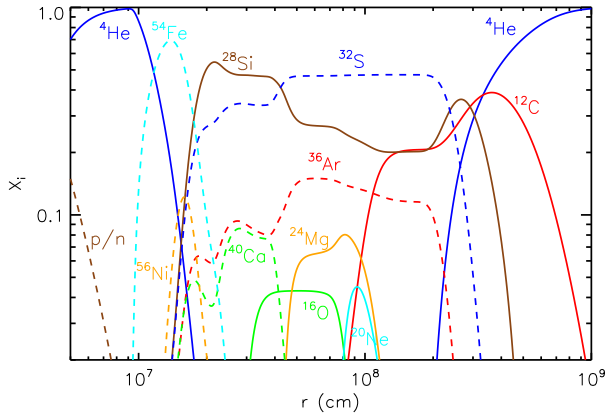}
}
\caption{Similar to Figure \ref{fig:CO}, but calculated for a pure He WD.  Values of $\eta_{\rm w} = 2$, Be$_{\rm d}^{'} = -0.1$, and $\alpha = 0.1$ are the same as the C-O case, but the latter corresponds to a lower accretion rate $\dot{M}(R_{\rm d}) \simeq 10^{-3}M_{\odot}$ s$^{-1}$ due to the larger outer radius $R_{\rm d} = 2\times 10^{9}$ cm.}
\label{fig:He}
\end{figure}

\begin{figure}
\centering
\subfigure[Radial Profile of Disk Properties]{
\includegraphics[width=0.46\textwidth]{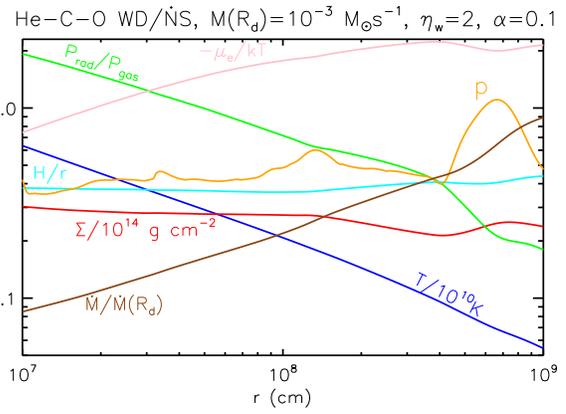}
}
\subfigure[Heating/Cooling Terms in the Entropy Equation]{
\includegraphics[width=0.46\textwidth]{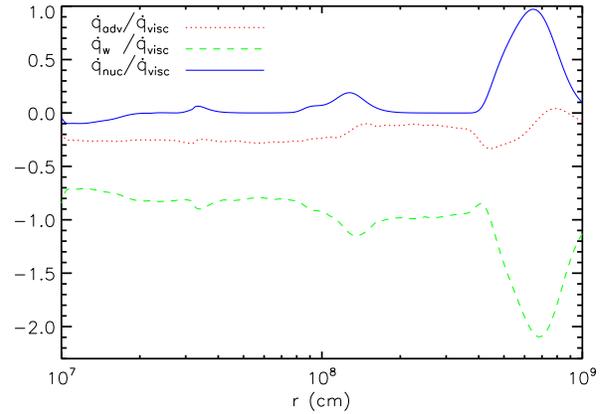}
}
\subfigure[Composition Profile]{
\includegraphics[width=0.46\textwidth]{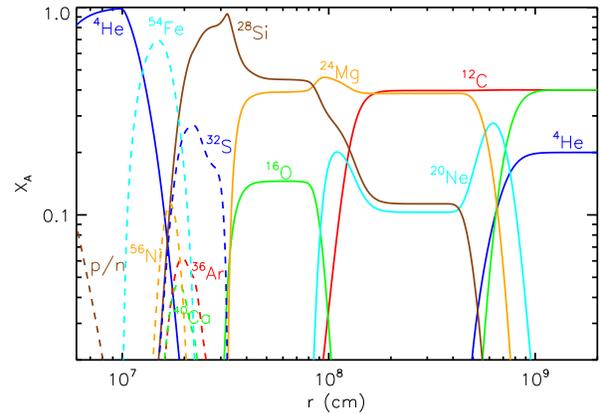}
}
\caption{Same as Figure \ref{fig:He}, but calculated for a `hybrid' He-C-O WD.}
\label{fig:hybrid}
\end{figure}

\begin{figure}
\centering
\subfigure[Radial Profile of Disk Properties]{
\includegraphics[width=0.46\textwidth]{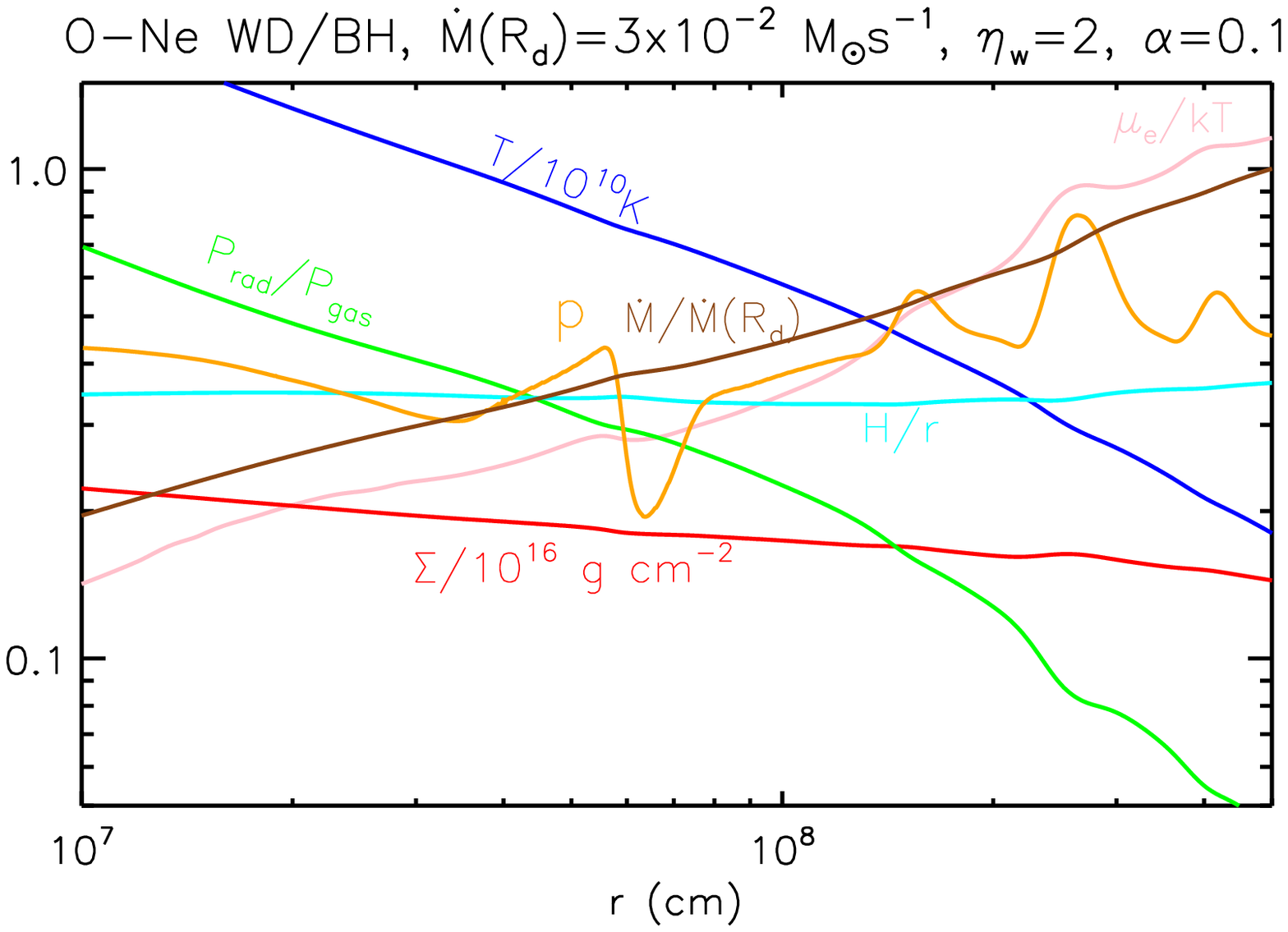}
}
\subfigure[Heating/Cooling Terms in the Entropy Equation]{
\includegraphics[width=0.46\textwidth]{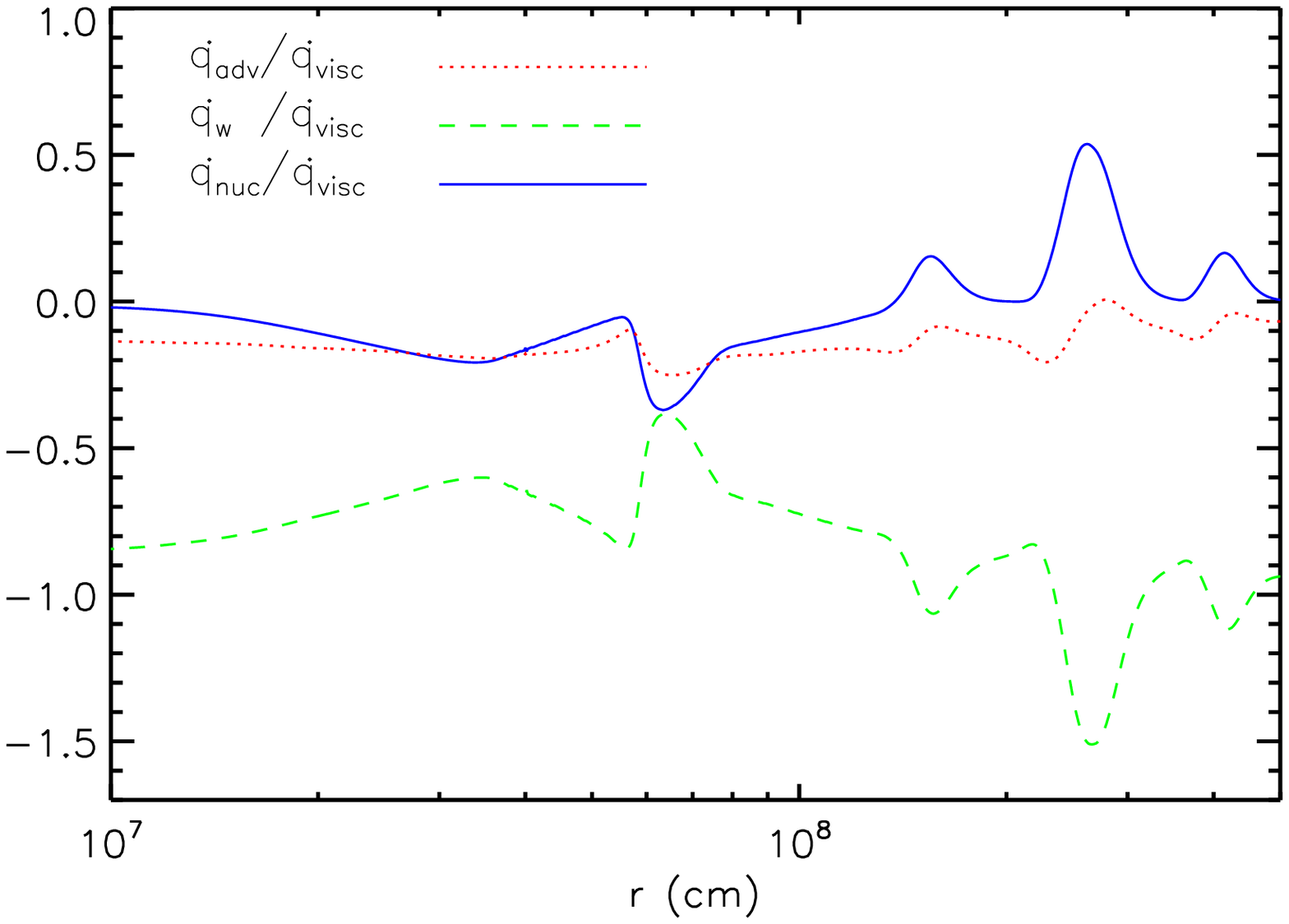}
}
\subfigure[Composition Profile]{
\includegraphics[width=0.46\textwidth]{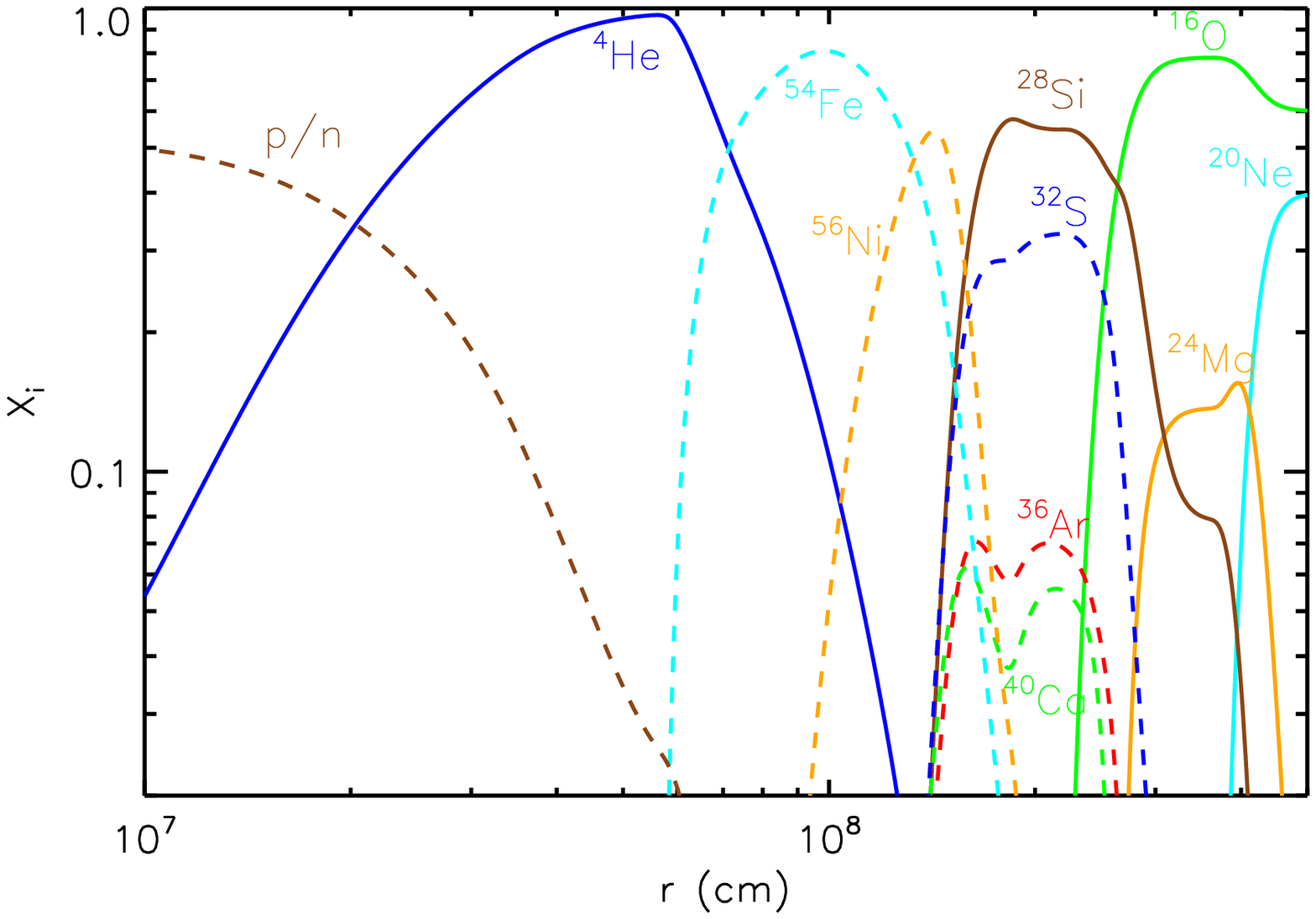}
}
\caption{Similar as Figures \ref{fig:CO}-\ref{fig:hybrid}, but calculated for the merger of a O-Ne WD with a 3$M_{\odot}$ BH.  We adopt values of $\eta_{\rm w} = 2$, Be$_{\rm d}^{'} = -0.2$, and $\alpha = 0.1$, and $\dot{M}(R_{\rm d}) \simeq 3\times 10^{-3}M_{\odot}$ s$^{-1}$, $R_{\rm d} = 5\times 10^{8}$ cm. }
\label{fig:ONe}
\end{figure}

Figure \ref{fig:CO} shows our baseline model for accretion following the merger of a $0.6M_{\odot}$ C-O WD with a 1.4$M_{\odot}$ NS.  In this example we adopt characteristic values for the initial accretion rate $\dot{M}(R_{\rm d}) = 3\times 10^{-3}M_{\sun}$ s$^{-1}$, viscosity $\alpha = 0.1$, disk Bernoulli parameter Be$_{\rm d}^{'} = -0.1$ (eq.~[\ref{eq:Bed}]), and wind strength $\eta_{w} = 2$.  The top panel shows the radial profiles of quantities characterizing the disk thermodynamics.  Note the following: (1) the disk is geometrically thick, with aspect ratio $H/r \approx 0.4$ at all radii, as expected for RIAFs; (2) gas pressure $P_{\rm gas}$ exceeds radiation pressure $P_{\rm rad}$, although they are comparable at small radii $\lesssim 10^{7}$ cm; (3) $\dot{M}$ decreases by a factor $\sim 10$ between $R_{\rm d} \approx 10^{9}$ cm and the inner radius $R_{\rm in} \approx 10^{7}$ cm, such that $\gtrsim 80\%$ of the accreting mass escapes in outflows.  This behavior is reflected in the wind mass loss index $p$ (eq.~[\ref{eq:p}]), which fluctuates with radius, depending on the local heating/cooling from nuclear reactions.

The middle panel in Figure \ref{fig:CO} shows sources of heating and cooling that contribute to the entropy equation (eq.~[\ref{eq:entropy}]), in ratio to the viscous heating rate $\dot{q}_{\rm visc}$ (eq.~[\ref{eq:qdotvisc}]).  At large radii ($r \gtrsim 5\times 10^{7}$ cm), both viscous dissipation and nuclear burning contribute to the heating, while cooling results from advection to smaller radii $\dot{q}_{\rm adv}$ and mass loss due to winds $\dot{q}_{\rm w}$ (eq.~[\ref{eq:qdotwind}]).  The nuclear heating rate achieves local maxima at radii where the temperature reaches the threshold necessary to burn the next element (e.g.~carbon burns at $r \approx 3\times 10^{8}$ cm and $T \approx 1.5\times 10^{9}$ K).  At these locations the energy generated by nuclear reactions is significant compared to that released by viscous dissipation; this general behavior has important implications for the thermal stability of the disk ($\S\ref{sec:stability}$).  At smaller radii ($r \lesssim 5\times 10^{7}$ cm), by contrast, $\dot{q}_{\rm nuc} < 0$ instead contributes to {\it cooling} the disk.  This occurs once temperatures are sufficiently high ($T \gtrsim 6\times 10^{9}$ K) to photo-disintegrate nuclei {\it endothermically}.  

The bottom panel of Figure \ref{fig:CO} shows the mass fractions of individual elements $X_{A}$ as a function of radius.  The composition at the outer edge of the disk is that of the initial WD: half $^{12}$C and half $^{16}$O.  Moving to smaller radii, carbon first burns to $^{20}$Ne and $^{4}$He, the latter of which rapidly capture onto $^{16}$O, eventually forming $^{24}$Mg.  At smaller radii and higher temperatures, $^{16}$O burns to $^{28}$Si.  Moving yet further in, photo-dissociation releases $\alpha-$particles, which immediately capture onto heavier nuclei (`Si burning'), producing $^{32}$S, $^{36}$Ar, $^{40}$Ca, $^{56}$Ni, and $^{54}$Fe.  Finally, at the smallest radii, $^{54}$Fe photo-disintegrates into $^{4}$He, which itself then disintegrates into free protons and neutrons.  

In Figure \ref{fig:He} we show similar results to Figure \ref{fig:CO}, but for the case of a 0.3$M_{\odot}$ pure He WD.  We adopt the same values for $\eta_{\rm w} = 2$, Be$_{\rm d}^{'} = -0.1$, and $\alpha = 0.1$, but the latter corresponds to a lower accretion rate $\dot{M}(R_{\rm d}) \simeq 10^{-3}M_{\odot}$ s$^{-1}$ than in the C-O WD case, due to the larger outer radius $R_{\rm d} = 2\times 10^{9}$ cm.  The qualitative disk structure is similar to the C-O case, but a few differences should be noted.  First, mass loss is considerably greater from the outer portions of the disk ($p \gtrsim 1$).  This is due to the substantial energy released by He burning and its low threshold temperature; burning thus occurs at large radii where the gravitational potential is shallow.  Mass loss greatly reduces the midplane density, which increases the importance of radiation pressure.  Because the temperature is lower in radiation-dominated than in gas pressure-dominated disks, burning heavier elements is delayed until smaller radii than in the C-O case.  Also note that the production of $^{16}$O, $^{20}$Ne and $^{24}$Mg are largely bypassed (e.g.~in comparison to the C-O case) because additional $\alpha$ captures are rapid following the onset of the [rate-limiting] triple-$\alpha$ reaction.

Figure \ref{fig:hybrid} shows results for a 0.4$M_{\odot}$ `hybrid' He-C-O WD, again for an accretion rate $\dot{M}(R_{\rm d}) = 10^{-3}M_{\odot}$ s$^{-1}$.  Although the inner structure of the disk resembles the C-O case (Fig.~\ref{fig:CO}), an important difference is again the onset of He burning at relatively large radii (low temperature), this time via the reaction $^{4}$He+$^{16}$O$\rightarrow ^{20}$Ne+$\gamma$.  As we discuss in $\S\ref{sec:stability}$, the sensitive temperature dependence of this reaction may render such disks thermally unstable.

Figure \ref{fig:ONe} shows results for a 1.2$M_{\odot}$ O-Ne WD accreting onto a 3$M_{\odot}$ BH.  The compact outer radius of the disk in this case $R_{\rm d} \simeq 5\times 10^{8}$ cm results in a high accretion rate $\dot{M}(R_{\rm d}) \simeq 3\times 10^{-2}M_{\odot}$ s$^{-1}$ for $\alpha = 0.1$.  Without He or C present, the first element to burn in this case is $^{20}$Ne at $T \approx 2\times 10^{9}$ K, but otherwise the composition profile is similar to the C-O case (Fig.~\ref{fig:CO}). 

In addition to the calculations presented above, we have explored the sensitivity of our results to variations in the binary parameters, WD composition, viscosity $\alpha$, and wind mass loss parameter $\eta_{w}$ (see Table \ref{table:results}).  In general, we find that increasing $\eta_{w}$ at fixed Be$_{\rm d}^{'}$ decreases the amount of mass loss because wind cooling is more effective.  By contrast, changing $\alpha$ (along with a compensating change in $\dot{M}(R_{\rm d})$; eq.~[\ref{eq:mdot_out}]) has a smaller effect.  Our results are also robust to realistic variations in the composition of the disrupted WD.    

Notably absent from the discussion thus far is the merger of a massive C-O or O-Ne WD with a NS companion.  The large mass ratio $q \sim 1$ in this case results in a compact disk ($R_{\rm d} \sim 10^{8}$ cm) with a very high initial accretion rate $\dot{M}(R_{\rm d}) \approx 0.1M_{\odot}$ s$^{-1}$ for $\alpha = 0.1$.  The initial temperature of the disk is sufficiently high that nuclear burning begins already at $r \approx R_{\rm d}$ during the circularization process itself.  Since virialization has not yet occurred, the WD material forming the disk may still be degenerate, potentially leading to an explosive situation.  Furthermore, as we discuss in $\S\ref{sec:stability}$, accretion in these systems may be thermally unstable, even under non-degenerate conditions.  Clearly, many of the simplifications adopted in our model break down in the case of massive WD-NS mergers.  Obtaining even a qualitative understanding of such events may require multi-dimensional simulations that include the effects of nuclear burning, even during the merger process itself. 

 \subsection{Outflow Properties}
\label{sec:outflowresults}

A schematic diagram of the outflows from WD-NS/BH merger disks is shown in Figure \ref{fig:cartoon}.  The total mass loss rate from the disk is approximately given by $\dot{M}_{\rm w} \simeq \dot{M}(R_{\rm d})-\dot{M}(R_{\rm in})$, where $R_{\rm in} \sim 10^{6.5}-10^{7.5}$ cm is the radius interior to which the disk is primarily free nucleons.  We adopt this inner boundary because if neutrino cooling is important at radii $r < R_{\rm in}$, then additional mass loss will be suppressed (although in $\S\ref{sec:SN}$ we discuss the possibility that the ejecta does contain a small fraction of free nucleons).  If most of the WD accretes at the characteristic rate $\dot{M}(R_{\rm d})$ (eq.~[\ref{eq:mdot_out}]), then the total ejecta mass is $M_{\rm ej} \approx f_{\rm ej}M_{\rm WD}$, where $f_{\rm ej} = \left[\frac{\dot{M}_{\rm w}}{\dot{M}_{\rm d}(R_{\rm d})}\right]$ is the fraction of the accreted mass lost in winds.  Depending on $\eta_{\rm w}$, we find that $f_{\rm w} \sim 0.5-0.99[0.90-0.99]$ for C-O[He] WDs (Table \ref{table:results}), corresponding to $M_{\rm ej} \sim 0.3-1.3[0.3]M_{\odot}$. 

The radial structure of the disk is imprinted on the composition of the outflows (Fig.~\ref{fig:cartoon}).  Table \ref{table:results} provides the mass fractions in each element of the wind ejecta, which are calculated as
\be
X_{\rm w,A} = \frac{1}{\dot{M}_{\rm w}}\int_{R_{\rm in}}^{R_{\rm d}}X_{A}\frac{\partial \dot{M}_{\rm w}}{\partial r}dr 
\label{eq:XA}
\ee
Here we have assumed that the local surface composition of the disk is similar to that in the midplane due e.g.~to turbulent mixing.  To compare with our C-O disk models, we also list the abundances predicted by the well-studied `W7' model for SN Ia from the delayed-detonation of a Chandrasekhar-mass C-O WD \citep{Nomoto+84}.  To compare with our He disk models, we list the range in ejecta abundances predicted from models for unstable He shell burning on the surface of a C-O WD, as calculated by \citet{Shen+10} for different values of the mass of the WD and He layer. 

In our fiducial C/O model (Fig.~\ref{fig:CO}) the majority of the ejecta is unburnt $^{12}$C ($26\%$) and $^{16}$O ($39\%$), with the remainder in $^{28}$Si ($10\%$), $^{24}$Mg ($7\%$), $^{4}$He ($6\%$), $^{20}$Ne ($5\%$), $^{54}$Fe ($3\%$), $^{32}$S($2\%$), $^{56}$Ni ($1.1\%$), $^{36}$Ar ($0.4\%$) and $^{40}$Ca ($0.3\%$).  Compared to the W7 SN Ia model, the fractional abundances of unburnt C/O, $^{20}$Ne, and $^{24}$Mg are significantly higher.  Despite these differences, the abundances of the elements Si, S, and Ca, generally responsible for producing the most prominent spectral features in normal Ia, are similar to the W7 model within a factor $\lesssim 2-3$.  Importantly, the yield of radioactive $^{56}$Ni is substantially lower than in normal Ia SN ejecta; this has important implications for the luminosities of optical transients associated with WD-NS/BH mergers ($\S\ref{sec:SN}$).

In our fiducial He model (Fig.~\ref{fig:He}) the vast majority of the ejecta is unburnt $^{4}$He ($63\%$), with the remainder in $^{12}$C ($15\%$), $^{28}$Si ($10\%$), $^{32}$S($8\%$), $^{36}$Ar ($2\%$), $^{54}$Fe ($0.8\%$), $^{24}$Mg ($0.6\%$),  $^{40}$Ca ($0.3\%$), $^{20}$Ne ($0.2\%$), $^{16}$O ($0.2\%$), and $^{56}$Ni ($0.1\%$).  Although the unburnt $^{4}$He fraction is similar to the He shell burning models of Shen et al.~(2010), the predicted abundances of the radioactive isotopes $^{44}$Ti, $^{48}$Cr, and $^{56}$Ni are all significantly lower, while those of intermediate mass elements are significantly higher.  In addition to the above elements, disk outflows may in all cases also contain a tiny fraction of $^{1}$H from the very inner disk, the implications of which are briefly discussed in $\S\ref{sec:SN}$.

Winds from smaller radii in the disk achieve higher velocities $v_{w} \propto v_{\rm k} \propto r^{-1/2}$ than those from further out.  Nevertheless, most of the ejecta will ultimately reside in a single `shell' of material with a much lower velocity dispersion.  This is because, initially after the disruption (on timescales $t \ll t_{\rm visc}$; eq.~[\ref{eq:tacc}]), the disk is concentrated at large radii $\approx R_{\rm d}$.  Slow outflows from this early stage will thus have time to encase the system before outflows have even begun in earnest from the inner disk on a timescale $t \gtrsim t_{\rm visc}$.  Fast ejecta from the inner disk will thus collide with the slower material ejected prior.  Because this interaction conserves energy (radiative losses are negligible at early times), the ejecta will achieve a mean velocity $\bar{v}_{\rm ej}$, which may be estimated by averaging over the total kinetic energy of winds from each disk annulus:
\be
\frac{1}{2}M_{\rm ej}\bar{v}_{\rm ej}^{2} = \frac{1}{2}\int_{R_{\rm in}}^{R_{\rm d}}\frac{\partial \dot{M}_{\rm w}}{\partial r}v_{w}^{2}dr 
\label{eq:vej}
\ee
Table \ref{table:results} shows that for C-O WD/NS mergers $\bar{v}_{\rm w}$ is typically $\sim 1-3\times 10^{4}$ km s$^{-1}$, while for He[O-Ne] mergers $v_{\rm w}$ is somewhat lower[higher].  

The reason that $\bar{v}_{\rm w}$ can exceed the typical velocity of SN Ia ejecta is the deeper gravitational potential well in the case of NS/BH accretion.  As a toy example, if mass loss were to occur with a constant power-law index $p = const < 1$ (eq.~[\ref{eq:p}]), then integrating equation (\ref{eq:vej}) gives 
\begin{eqnarray}
\bar{v}_{\rm ej} \simeq 
\left\{
\begin{array}{lr}
 \left[\frac{p\eta_{\rm w}}{1-p}\frac{GM}{R_{\rm in}}\left(\frac{R_{\rm in}}{R_{\rm d}}\right)^{p}\right]^{1/2}
, &
p < 1 \\
 \left[\eta_{\rm w}\frac{GM}{R_{\rm d}}{\rm ln}\left(\frac{R_{\rm d}}{R_{\rm in}}\right)\right]^{1/2}
, &
p = 1 \\
\end{array}
\right., \nonumber \\
\label{eq:vej2}
\end{eqnarray}
where in the first expression we have assumed that $R_{\rm d} \gg R_{\rm in}$.  A larger wind efficiency factor $\eta_{\rm w}$ thus increases $\bar{v}_{\rm ej}$, both because $v_{\rm w} \propto \eta_{w}^{1/2}$ and because larger $\eta_{\rm w}$ decreases the mass loss rate $p(r)$ required to cool the disk.  Equation (\ref{eq:vej2}) shows that (for $\eta_{\rm w} \simeq 1$) the minimum value of $v_{\rm ej}$ ($p \approx 1$) is of the order of the orbital velocity at $r \approx R_{\rm d}$.  This varies from $\approx 3\times 10^{3}$ km s$^{-1}$ for He WD-NS mergers to $\approx 10^{4}$ km s$^{-1}$ for O-Ne WD-NS/BH mergers.

\begin{table*}
\begin{center}
\vspace{0.05 in}\caption{Properties of Disk Models}
\label{table:models}
\resizebox{17cm}{!}
{
\begin{tabular}{lccccccccccc}
\hline
\hline
\multicolumn{1}{c}{model} &
\multicolumn{1}{c}{$M^{(a)}$} &
\multicolumn{1}{c}{$M_{\rm WD}$} &
\multicolumn{1}{c}{$R_{\rm d}^{(b)}$} &
\multicolumn{1}{c}{$\dot{M}(R_{\rm d})$} &
\multicolumn{1}{c}{$\alpha$} &
\multicolumn{1}{c}{Be$_{\rm d}^{'(c)}$} &
\multicolumn{1}{c}{$\eta_{\rm w}^{(d)}$} &
\multicolumn{1}{c}{$X_{A}(R_{\rm d})^{(e)}$} &
\multicolumn{1}{c}{$\frac{\dot{M}_{\rm wind}}{\dot{M}(R_{\rm d})}^{(f)}$} &
\multicolumn{1}{c}{$\bar{v}_{\rm ej}^{(g)}$} &
\multicolumn{1}{c}{unstable?$^{(h)}$} 
\\
& ($M_{\sun}$) & ($M_{\sun}$) & (cm) & ($M_{\sun}$ s$^{-1}$) & & & & & & ($10^{4}$ km s$^{-1}$) & \\
\hline 
\\

NS\_C-O\_1 & 1.4 & 0.6 & $10^{9}$ & $3\times 10^{-3}$ & 0.1 & -0.1 & 2 & $X_{12} = 0.5, X_{16} = 0.5$ & 0.84 & 2.8 & no \\ 

NS\_C-O\_2 & 1.4 & 0.6 & $10^{9}$ & $3\times 10^{-4}$ & 0.01 & -0.1 & 2 & - & 0.86& 2.8 & no \\ 

NS\_C-O\_3 & 1.4 & 0.6 & $10^{9}$ & $3\times 10^{-3}$ & 0.1 & -0.2 & 1 & - & 0.99 & 1.6 & no \\ 

NS\_C-O\_4 & 1.4 & 0.6 & $10^{9}$ & $3\times 10^{-3}$ & 0.1 & 0.0 & 3 & - & 0.57 & 3.2 & no \\

NS\_C-O-He\_1 & 1.4 & 0.4 & $2\times 10^{9}$ & $10^{-3}$ & 0.1 & -0.1 & 2 & $X_{16}=0.4,X_{12}=0.4,X_{4}=0.2$ & 0.95 & 2.5 & yes \\

NS\_C-O-He\_2 & 1.4 & 0.4 & $2\times 10^{9}$ & $10^{-4}$ & 0.01 & -0.1 & 2 & - & 0.95 & 2.4 & yes \\

NS\_O-Ne$^{(i)}$ & 1.4 & 1.2 & $2\times 10^{8}$ & $10^{-1}$ & 0.1 & - & - & $X_{16} = 0.6, X_{20} = 0.4$ & - & - & yes \\

BH\_C-O\_1 & 3 & 1.0 & $5\times 10^{8}$ & $3\times 10^{-2}$ & 0.1 & -0.2 & 2 & $X_{12} = 0.5, X_{16} = 0.5$ & 0.72 & 3.6 & yes \\

BH\_C-O\_2 & 3 & 1.0 & $5\times 10^{8}$ & $3\times 10^{-3}$ & 0.01 & -0.2 & 2 & check & 0.72 & 3.6 & yes \\

BH\_C-O\_3 & 3 & 1.0 & $5\times 10^{8}$ & $3\times 10^{-2}$ & 0.1 & -0.25 & 1 & - & 0.96 & 2.3 & yes \\

BH\_O-Ne\_1 & 3 & 1.2 & $5\times 10^{8}$ & $3\times 10^{-2}$ & 0.1 & -0.2 & 2 & $X_{16} = 0.6, X_{20} = 0.4$ & 0.83 & 4.6 & yes \\

BH\_O-Ne\_2 & 3 & 1.2 & $5\times 10^{8}$ & $3\times 10^{-3}$ & 0.01 & -0.2 & 2 & - & 0.72 & 3.5 & yes \\

NS\_He\_1 & 1.2 & 0.3 & $2\times 10^{9}$ & $10^{-3}$ & 0.1 & -0.1 & 2 & $X_{4} = 1$ & 0.95 & 2.2 & no \\

NS\_He\_2 & 1.2 & 0.3 & $2\times 10^{9}$ & $10^{-4}$ & 0.01 & -0.1 & 2 & - & 0.96 & 2.0 & no \\

NS\_He\_3 & 1.2 & 0.3 & $2\times 10^{9}$ & $10^{-3}$ & 0.1 & -0.2 & 1 & - & 0.99 & 1.0 & no \\
\hline
\end{tabular}
}
\end{center}
{\small $^{(a)}$ Mass of the central NS or BH.  $^{(b)}$Outer edge of the disk $R_{\rm d} = R_{\rm RLOF}(1+q)^{-3}$, where $q \equiv M_{\rm WD}/M$ and $R_{\rm RLOF}$ is the binary separation at RLOF (eq.~[\ref{eq:rRLOF}]).  $^{(c)}$Normalized Bernoulli parameter in the disk (eq~.[\ref{eq:Bed}]).  $^{(d)}$ Ratio of the specific kinetic energy of the wind from any radius to the local gravitational binding energy.  $^{(e)}$Initial composition of the disk from the disrupted WD.  $^{(f)}$Ratio of the total mass loss rate in the wind $\dot{M}_{\rm w}\equiv \dot{M}(R_{\rm d})-\dot{M}(R_{\rm in})$ to the initial accretion rate $\dot{M}(R_{\rm d})$ at the outer edge of the disk. $^{(g)}$Mean velocity of the ejecta from disk winds (eq.~[\ref{eq:vej}]).   $^{(h)}$Whether the disk is thermally unstable due to nuclear burning at any radius, according to the most stringent criterion $f_{\rm nuc} < f_{\rm nuc,th2}$ given in equation (\ref{eq:thermal}).  $^{(i)}$Characteristic values shown for illustration only.  The disk model presented in this paper cannot be applied to the merger of massive WD-NS binaries because nuclear burning begins already at $r \approx R_{d}$ (see $\S\ref{sec:outflowresults}$). }
\end{table*}

\begin{table*}
\begin{center}
\vspace{0.05 in}\caption{Elemental Mass Fractions in the Wind Ejecta $X_{\rm w,A}$ (eq.~[\ref{eq:XA}])}
\label{table:results}
\resizebox{18cm}{!}
{
\begin{tabular}{lcccccccccccccc}
\hline
\hline
\multicolumn{1}{c}{model} &
\multicolumn{1}{c}{$^{4}$He} &
\multicolumn{1}{c}{$^{12}$C} &
\multicolumn{1}{c}{$^{16}$O} &
\multicolumn{1}{c}{$^{20}$Ne} &
\multicolumn{1}{c}{$^{24}$Mg} &
\multicolumn{1}{c}{$^{28}$Si} &
\multicolumn{1}{c}{$^{32}$S} &
\multicolumn{1}{c}{$^{36}$Ar} &
\multicolumn{1}{c}{$^{40}$Ca} &
\multicolumn{1}{c}{$^{44}$Ti} &
\multicolumn{1}{c}{$^{48}$Cr} &
\multicolumn{1}{c}{$^{52}$Fe} &
\multicolumn{1}{c}{$^{54}$Fe} &
\multicolumn{1}{c}{$^{56}$Ni} 
\\
 & & & & & & & & & & &  & & \\
\hline 
\\
W7 Ia$^{(a)}$ & 0.0 & 0.036 & 0.10 & 0.0014 & 0.0094 & 0.11 & 0.060 & 0.11 & 0.0086 & $5.7\times 10^{-6}$ & $3.5\times 10^{-9}$ & - & 0.054 & 0.42 \\

NS\_C-O\_1& 0.063 & 0.26 & 0.39 & 0.049 & 0.065 & 0.10 & 0.023 & 0.0043 & 0.0031 & $2.1\times 10^{-5}$ & $1.0\times 10^{-4}$ & $9.7\times 10^{-4}$ & 0.032 & 0.011 \\

NS\_C-O\_2 & 0.052 & 0.26 & 0.38 & 0.058 & 0.067 & 0.096 & 0.021 & 0.0036 & 0.0025 & $1.3\times 10^{-5}$ & $7.0\times 10^{-5}$ & $8.0\times 10^{-4}$ & 0.042 & 0.011 \\

NS\_C-O\_3 & 0.0091 & 0.41 & 0.47 & 0.029 & 0.034 & 0.034 & 0.0068 & 0.0012 & $8.0\times 10^{-4}$ & $3.7\times 10^{-6}$ & $1.4\times 10^{-5}$ & $1.0\times 10^{-4}$ & 0.0061 & $1.0\times 10^{-3}$ \\

NS\_C-O\_4 & 0.070 & 0.21 & 0.36 & 0.057 & 0.076 & 0.13 & 0.029 & 0.0055 & 0.0040 & $2.6\times 10^{-5}$ & $1.3\times 10^{-4}$ & 0.0012 & 0.035 & 0.014 \\

NS\_C-O-He\_1 & 0.12 & 0.33 & 0.18 & 0.093 & 0.16 & 0.087 & 0.0064 & 0.0010 & $6.5\times 10^{-4}$ & $4.2\times 10^{-6}$ & $1.6\times 10^{-5}$ & $1.2\times 10^{-4}$ & 0.010 & 0.0010 \\

NS\_C-O-He\_2  & 0.11 & 0.33 & 0.17 & 0.095 & 0.19 & 0.071 & 0.0054 & 9.5$\times 10^{-4}$ & $7.1\times 10^{-4}$ & $4.5\times 10^{-6}$ & $2.3\times 10^{-5}$ & $2.3\times 10^{-4}$ & 0.011 & 0.0027 \\

BH\_C-O\_1  & 0.15 & $6.7\times 10^{-4}$ & 0.20 & 0.043 & 0.11 & 0.24 & 0.047 & 0.0081 & 0.0063 & $4.8\times 10^{-5}$ & 2.9$\times 10^{-4}$ & 0.0036 & 0.12 & 0.053 \\ 

BH\_C-O\_2 & 0.14 & $5.7\times 10^{-5}$ & 0.19 & 0.021 & 0.098 & 0.25 & 0.047 & 0.0062 & 0.0039 & $1.2\times 10^{-5}$ & $6.3\times 10^{-5}$ & $7.4\times 10^{-4}$ & 0.22 & 0.011 \\ 

BH\_C-O\_3 & 0.062 & 0.12 & 0.30 & 0.10 & 0.098 & 0.17 & 0.032 & 0.0053 & 0.0040 & $3.0\times 10^{-5}$ & $1.8\times 10^{-4}$ & 0.0021 & 0.065 & 0.031 \\ 

BH\_O-Ne\_1 & 0.15 & $4.7\times 10^{-6}$ & 0.30 & 0.045 & 0.035 & 0.17 & 0.076 & 0.016 & 0.013 & $5.5\times 10^{-5}$ & $3.0\times 10^{-4}$ & 0.0036 & 0.12 & 0.054 \\ 

BH\_O-Ne\_2  & 0.13 & $2.8\times 10^{-6}$ & 0.27 & 0.024 & 0.037 & 0.18 & 0.080 & 0.015 & 0.011 & $1.8\times 10^{-5}$ & $6.6\times 10^{-5}$ & $7.1\times 10^{-4}$ & 0.22 & 0.010 \\

Shen10 .Ia$^{(b)}$ & $0.30-0.45$ & - & - & - & - & - & - & $6-700\times 10^{-5}$ & $0.001-0.04$ & $0.007-0.1$ & $0.01-0.1$ & $0.03-0.1$ & - & 0.1-0.7 \\

NS\_He\_1 & 0.63 & 0.15 & 0.0022 & 0.0017 & 0.0055 & 0.095 & 0.082 & 0.018 & 0.0027 & $7.1\times 10^{-6}$ & $1.8\times 10^{-5}$ & $1.1\times 10^{-4}$ & 0.0082 & $9.7\times 10^{-4}$ \\ 

NS\_He\_2 & 0.60 & 0.23 & 0.0031 & 0.0036 & 0.016 & 0.13 & 0.013 & $7.8\times 10^{-4}$ & 4.3$\times 10^{-4}$ & $2.7\times 10^{-6}$ & $1.4\times 10^{-5}$ & $1.4\times 10^{-4}$ & 0.0068 & $1.6\times 10^{-3}$ \\ 

NS\_He\_3 & 0.87 & 0.077 & $4.3\times 10^{-4}$ & $4.7\times 10^{-4}$ & 0.0014 & 0.024 & 0.024 & 0.0092 & 7.7$\times 10^{-4}$ & $1.4\times 10^{-6}$ & $1.0\times 10^{-6}$ & $1.8\times 10^{-6}$ & $4.0\times 10^{-4}$ & $5.3\times 10^{-6}$ \\ 

\hline
\end{tabular}
}
\end{center}
{\small $^{(a)}$W7 Type Ia SN model \citep{Nomoto+84}, $^{(b)}$Range of models for unstable He shell burning from \citet{Shen+10}.}
\end{table*}

\subsection{Thermal Stability}
\label{sec:stability}

\begin{figure}
\centering
\subfigure[0.6$M_{\odot}$ WD/NS]{
\includegraphics[width=0.48\textwidth]{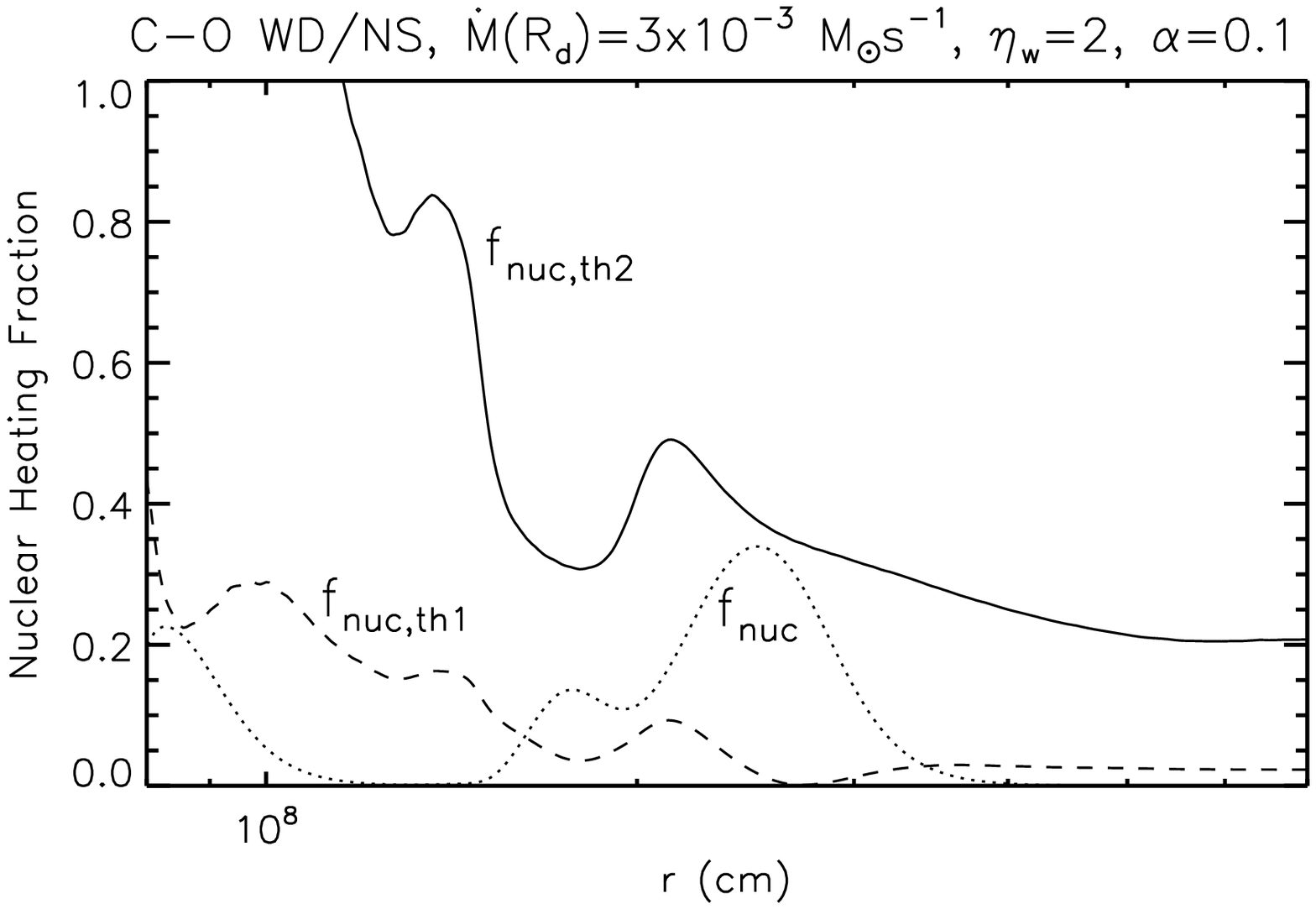}
}
\subfigure[0.4$M_{\odot}$ He-C-O/NS, 1.0$M_{\odot}$ WD/BH, and 1.2$M_{\odot}$ WD/BH]{
\includegraphics[width=0.48\textwidth]{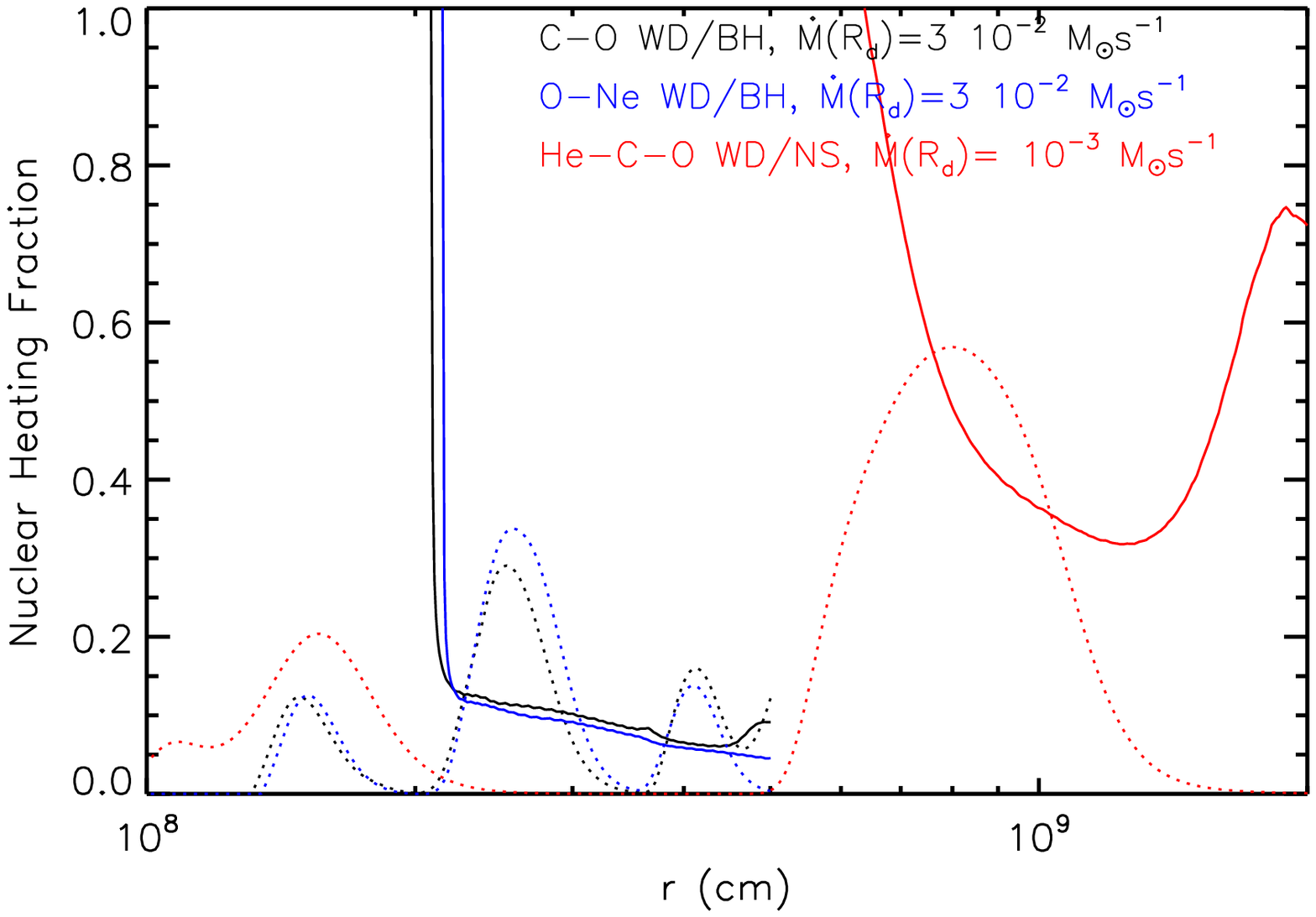}
}
\caption{Analysis of the thermal stability of the accretion solutions presented in $\S\ref{sec:diskresults}$.  Quantities shown are the fraction of the total heating due to nuclear burning $f_{\rm nuc} \equiv \dot{q}_{\rm nuc}/\dot{q}^{+}$ ({\it dotted line}) and the threshold fractions for thermal instability $f_{\rm nuc,th1}$ ({\it dashed line}) and $f_{\rm nuc,th2}$ ({\it solid line}) under the assumption that $v_{\rm w} \propto v_{k}$ and $v_{\rm w} \propto a$, respectively (eq.~[\ref{eq:fnucth}]).  {\bf Top }: Solutions corresponding to accretion following the merger of a 0.6$M_{\odot}$ 1.4$M_{\odot}$ WD with a NS at the rate $\dot{M}(R_{\rm d}) = 3\times 10^{-3}M_{\odot}$ s$^{-1}$ (Fig.~\ref{fig:CO}).  Note that although $f_{\rm nuc}<f_{\rm nuc,th2}$ at all radii, $f_{\rm nuc}$ exceeds $f_{\rm nuc,th1}$ across some range in radius.  This shows that whether the disk is in fact stable depends on the detailed pressure-dependence of wind cooling.  {\bf Bottom}:  Accretion following the merger of a $1M_{\odot}$ C-O WD with a 3$M_{\odot}$ BH ({\it Black}); $1.2M_{\odot}$ O-Ne WD with a 3$M_{\odot}$ BH ({\it Blue}); `hybrid' He-C-O WD with a 1.4$M_{\odot}$ NS ({\it Red}).  In all three cases $f_{\rm nuc} > f_{\rm nuc,th2}$ over some range in radii.  The solutions are thus unstable to runaway heating from nuclear burning, regardless of whether $v_{\rm w} \propto v_{k}$ or $v_{\rm w} \propto a$.}
\label{fig:stability}
\end{figure}

The calculations presented in the previous sections assumed steady-state accretion.  This is not valid, however, if the disk is thermally unstable, which might be expected due to the sensitive temperature dependencies of nuclear reactions.  The criterion for thermal stability may be written \citep{Piran78}
\be
\left.\frac{\partial{\rm ln}q^{+}}{\partial{\rm ln}H}\right|_{\rm \Sigma} < \left.\frac{\partial{\rm ln}q^{-}}{\partial{\rm ln}H}\right|_{\rm \Sigma}\rightarrow\,\,\,\left.\frac{\partial{\rm ln}q^{+}}{\partial{\rm ln}P}\right|_{\rm \Sigma} < \left.\frac{\partial{\rm ln}q^{-}}{\partial{\rm ln}P}\right|_{\rm \Sigma}
\label{eq:thermal}
\ee
where $q^{+} = \dot{q}_{\rm visc} + \dot{q}_{\rm nuc}$ and $\dot{q}^{-} = \dot{q}_{\rm adv} + \dot{q}_{\rm w}$ are the total heating and cooling rates, respectively (see eq.~[\ref{eq:entropy}]), and the second condition follows because $P = \rho a^{2} \propto \Omega_{\rm k}^{2}\Sigma H$.  Viscous heating obeys $\dot{q}_{\rm visc} \propto P^{2}$, while advective cooling obeys $\dot{q}_{\rm adv} \propto \dot{q}_{\rm visc}(H/r)^{2} \propto P^{4}$.  In our model, we have assumed that the terminal speed of disk winds $v_{\rm w}$ is proportional to the local escape speed $\propto v_{\rm k}$ (eq.~[\ref{eq:qdotwind}]), such that $\dot{q}_{\rm w} \propto \dot{q}_{\rm visc} \propto P^{2}$.  However, the precise mechanism responsible for driving the wind is uncertain; we could equally well have assumed that $v_{\rm w}$ is proportional to the sound speed $a$, in which case $\dot{q}_{\rm w} \propto (H/r)^{2}v_{\rm k} \propto P^{4}$.  Considering both cases, the stability criterion in equation (\ref{eq:thermal}) can equivalently be written as an upper limit on the fraction $f_{\rm nuc} \equiv \dot{q}_{\rm nuc}/\dot{q}^{+}$ of the disk heating supplied by nuclear reactions, viz.~
\begin{eqnarray}
f_{\rm nuc} < \left\{
\begin{array}{lr}
  \frac{2(1-f_{\rm w})}{n-2} \equiv f_{\rm nuc,th1}
, &
v_{\rm w} \propto v_{\rm esc} \\
 \frac{2}{n-2} \equiv f_{\rm nuc,th2}
,&
v_{\rm w} \propto a, \\
\end{array}
\right. \nonumber \\
\label{eq:fnucth}
\end{eqnarray}
where $n \equiv \left(\frac{\partial{\rm ln}\dot{q}_{\rm nuc}}{\partial {\rm ln}P}\right)|_{\Sigma}$ and $f_{\rm w} \equiv \dot{q}_{\rm w}/\dot{q}^{-}$ is the fraction of the cooling resulting from winds.  Equation (\ref{eq:fnucth}) shows that accretion is stable at a given radius if either (1) nuclear burning is absent entirely; or (2) the burning rate has a weak pressure dependence $n < 4-2f_{\rm w}$ or $n < 4$, in the cases $v_{\rm w} \propto v_{\rm k}$ and $v_{\rm w} \propto a$, respectively.  

Figure \ref{fig:stability} shows $f_{\rm nuc}$ ({\it solid line}), $f_{\rm nuc,th1}$ ({\it solid lines}), and $f_{\rm nuc,th2}$ as a function of radius, calculated for our solution from $\S\ref{sec:results}$ corresponding to accretion following a 0.6$M_{\odot}$ C-O WD (Fig.~\ref{fig:CO}).  Figure \ref{fig:stability} shows that in the case $v_{\rm w} \propto a$, the disk is stable at all radii, i.e.~$f_{\rm nuc} < f_{\rm nuc,th2}$.  If, however, we instead assume that $v_{\rm w} \propto v_{\rm k}$, then we find that the disk is unstable ($f_{\rm nuc} > f_{\rm th1}$) at the radii $\sim$ few $\times 10^{8}$ cm where carbon burning peaks.  Although we do not show this case explicitly, we find a simlilar result for He WDs: the triple-$\alpha$ reaction is not sufficiently pressure-sensitive for instability if $v_{\rm w} \propto a$ (i.e.~$n < 4$), but for $v_{\rm w} \propto v_{\rm k}$ the disk is unstable at radii $\sim 3-4\times 10^{8}$ cm where He burning peaks.  We conclude that whether C-O or He WD disks are thermally stable depends on the pressure dependence of wind cooling, which in turn depends on the [uncertain] mechanism responsible for driving the wind. 

In the case of O-Ne and hybrid WDs, thermal instability seems more assured.  In the bottom panel of Figure \ref{fig:stability} we show $f_{\rm nuc}$ and $f_{\rm nuc,th2}$ (eq.~[\ref{eq:fnucth}]) for cases corresponding to the merger of a 0.4$M_{\odot}$ He-C-O WD with a NS (Fig.~\ref{fig:hybrid}); a $1.2M_{\odot}$ O-Ne WD with a 3$M_{\odot}$ BH (Fig.~\ref{fig:ONe}); and a $1.0M_{\odot}$ C-O WD with a 3$M_{\odot}$ BH.  In all three cases we find that over some range of radii, $f_{\rm nuc}$ exceeds even the more conservative threshold for instability $f_{\rm nuc,th2}$.  This implies that our calculated solutions are thermally unstable, indpendent of whether $v_{\rm w} \propto v_{\rm k}$ or $v_{\rm w} \propto a$.  
Although addressing the implications of unstable burning is beyond the scope of this paper, we speculate that the end result may be a complicated, time-dependent evolution.  One possibility is `limit-cycle' behavior, in which the initial inflow is halted by runaway burning, before resuming again later with fresh fuel (a possible analog are dwarf nova outbursts of cataclysmic variables; e.g.~\citealt{Cannizzo93}).  Though it is not obvious, the steady-state solutions constructed here may remain a reasonable description of the average flow over many cycles.

\subsection{Effects of Convection}
\label{sec:convection}

It is well known that ADAFs are unstable to radial convection (\citealt{Narayan&Yi94}; \citealt{Igumenshchev+00}; \citealt{Quataert&Gruzinov00}).  In the case of NuDAFs, the disk may also be unstable to {\it vertical} convection because the nuclear heating peaks sharply near the midplane due to its sensitive temperature dependence.  Although our calculations have thus far neglected its effects on the transport of angular momentum, energy, and composition, in this section we describe how strong convection could alter our conclusions.  

If convection transports angular momentum outwards$-$similar to MHD turbulence produced by the MRI$-$then it would act simply to enhance the effective value of ``$\alpha$'', which we have shown does not qualitatively affect our results.  If, on the other hand, convection transports angular momentum inwards \citep{Ryu&Goodman92}, then an qualitatively different type of solution could obtain: a ``convection dominated accretion flow'' (CDAF; \citealt{Quataert&Gruzinov00}; \citealt{Narayan+00}).  In CDAFs, the angular momentum transported outwards by viscosity is balanced by the inward transport by convection, such that the net mass accretion rate is very small.  Because in steady state the convective energy flux $F_{c} \propto \rho c_{s}^{3}r^{2}$ is constant with radius, CDAFs are characterized by a more shallow density profile $\rho \propto r^{-1/2}$ than for normal ADAF solutions ($\rho \propto r^{-3/2}$).  If present, the CDAF would extend from the inner edge of the RIAF at $R_{\rm in} \sim 10^{6}-10^{7}$ cm to much larger radii, where the outflowing energy is released through radiation or outflows.   

Coincidentally, the density profile of a CDAF scales the same way with radius as in ADIOS (wind) models with ``maximal'' mass loss ($p = 1$; $\dot{M} \propto r$).  Our steady-state calculations with high mass loss rates (low $\eta_{w}$) thus may also be applied to the CDAF case.  Note, however, that the physical interpretation is completely different: in wind models, the outflowing mass escapes the system entirely, while in CDAFs it remains bound and is simply `recycled' by mixing to large radii, from which it accretes again later.  

If the outer edge of the disk remains near the initial radius of the disrupted WD $R_{\rm d}$, then convection simply increases the timescale for matter to accrete from that given in equation (\ref{eq:tacc}) by a factor $\sim \frac{\dot{M}(R_{\rm d})}{\dot{M}(R_{\rm in})} \sim \frac{R_{\rm d}}{R_{\rm in}} \sim 10-10^{3}$, or from minutes to hours or days.  The radial structure of the disk is similar to the maximal wind case, but the observational signature of the event as described in $\S\ref{sec:observe}$ will be altered.  With no outflows present, the central NS or BH accretes significantly more mass, potentially resulting in more energetic high energy emission ($\S\ref{sec:GRB}$).  On the other hand, with little or no radioactive material ejected, the SN-like optical counterpart (if any) would be much dimmer ($\S\ref{sec:SN}$).  Outflows may still occur from the outer radius of the CDAF in this scenario, but the mean velocity is likely be much lower and the $^{56}$Ni fraction would be smaller.

A more radical possibility is that the outward convective energy flux will `feed back' on the outer disk, alterating its dynamics entirely.  If pressure forces cause the outer reservoir of mass to expand significantly from its initial radius $\sim R_{\rm d}$, then rotational support will no longer be important.  As a much larger, quasi-spherical accreting envelope, the structure may thus come to resemble a similar `Thorne-Zytkow object' \citep{Thorne&Zytkow75} or `quasi-star' (e.g.~\citealt{Begelman+08}).  In this case, the accretion timescale will be even long than estimated above, because it instead depends on the rate that the convective energy escapes the outer boundary in radiation or winds.  A quantitative exploration of such a model is an important and interesting excercise for future work, but is beyond the scope of this paper.

Despite the important possible effects of convection discussed above, it is not clear that CDAFs are in fact relevant to WD-NS/BH mergers.  As already noted, whether the flow is best described as a CDAF or ADAF (with winds) depends on the strength (and direction) of convective angular momentum transport relative to other torques.  Although radial convection in hydrodynamic simulations of RIAFs indeed appears to balance outwards viscous transport for $\alpha \lesssim 0.05$ (Naryan et al.~2000; Narayan et al.~2004), higher values of $\alpha \sim 0.1$ are generally inferred from observations of ionized disks in a variety of astrophysical contexts \citep{King+07b}.  In WD-NS/BH mergers, angular momentum transport due to gravitational instabilities may also compete with convection.   Finally, it has been argued that global MHD simulations of RIAFs more closely  resemble the ADIOS picture of outflows employed in this paper (\citealt{Hawley+01}; \citealt{Hawley&Balbus02}), although this conclusion may depend on the [uncertain] saturation strength of the MRI and the ultimate fate of the mass leaving the grid.

\subsection{Analytic Estimate of the Importance of Nuclear Burning}
\label{sec:analytic}

We conclude this section with an analytic estimate of the importance of nuclear burning in hyper-accreting disks, i.e. we address when a disk is in the NuDAF regime.  Focusing on the burning of one isotope $A$, the nuclear heating rate may be written $\dot{q}_{\rm nuc} = \Delta\epsilon_{\rm nuc}/m_n t_{\rm burn}$, where $\Delta\epsilon_{\rm nuc} = X_{A}Q/A$ is the nuclear energy released per nucleon, $Q$ is the Q-value of the reaction, $m_n$ is the nucleon mass, $t_{\rm burn} = \Delta r_{\rm burn}/v_{\rm r}(r_{\rm burn})$ is the time spent burning at radius $r = r_{\rm burn}$, $v_{r} = 3\nu/2r$ is the accretion velocity, and $\Delta r_{\rm burn}$ is the radial distance over which most of the energy is released.  The ratio of nuclear to viscous heating can thus be written
\be
\left.\frac{\dot{q}_{\rm nuc}}{\dot{q}_{\rm visc}}\right|_{r=r_{\rm burn}} = \left.\frac{\Delta\epsilon_{\rm nuc}/m_n t_{\rm burn}}{(9/4)\nu\Omega_{k}^{2}}\right|_{r = r_{\rm burn}} = \frac{2}{3}\frac{\Delta\epsilon_{\rm nuc}r_{\rm burn}}{GMm_n}\frac{r_{\rm burn}}{\Delta r_{\rm burn}}    
\ee 
where we have used equation (\ref{eq:qdotvisc}), neglecting terms $\propto (H/r)^{2}$.  

Due to the temperature sensitivity of nuclear burning rates, $r = r_{\rm burn}$ generally occurs within a factor $\lesssim 2$ of a fixed temperature $T = T_{\rm burn}$, which depends primarily just on the element under consideration.  Because the radius and temperature of burning are related by
\begin{eqnarray}
r_{\rm burn} \simeq 
\left\{
\begin{array}{lr}
 \frac{\mu GMm_n}{kT_{\rm burn}}\left(\frac{H}{r}\right)^{2}
, &
P_{\rm gas} \gg P_{\rm rad} \\
 \left(\frac{\dot{M}G^{1/2}M^{1/2}}{2\pi\alpha a T_{\rm burn}^{4}}\frac{r}{H}\right)^{2/5}
, &
P_{\rm rad} \gg P_{\rm gas}, \\
\end{array}
\right. \nonumber \\
\label{eq:rburn}
\end{eqnarray}
in gas and radiation-dominated regimes, respectively, we can write
\begin{eqnarray}
&&\left.\frac{\dot{q}_{\rm nuc}}{\dot{q}_{\rm visc}}\right|_{r=r_{\rm burn}} = \frac{H}{\Delta r_{\rm burn}}\times \nonumber \\
&&\left\{
\begin{array}{lr}
 7.5\left(\frac{\Delta\epsilon_{\rm nuc}}{\rm MeV}\right)\left(\frac{T_{\rm burn}}{10^{9}{\,\rm K}}\right)^{-1}
, &
\frac{P_{\rm gas}}{P_{\rm rad}} \gg 1 \\
3\left(\frac{\Delta\epsilon_{\rm nuc}}{\rm MeV}\right)\left(\frac{M}{1.4M_{\odot}}\right)^{-0.8}\left(\frac{\alpha}{0.1}\right)^{-0.4}\left(\frac{\dot{M}}{10^{-3}M_{\odot}{\rm\,s^{-1}}}\right)^{0.4}\left(\frac{T_{\rm burn}}{10^{9}{\,\rm K}}\right)^{-1.6},&
\frac{P_{\rm rad}}{P_{\rm gas}} \gg 1, \\
\end{array}
\right.. \nonumber \\
\label{eq:qdotratio}
\end{eqnarray}
where we have assumed that $H/r = 0.4$ and a mean molecular weight $\mu = 2$.  Assuming nominal gas pressure dominance, the ratio of radiation to gas pressure is given by
\begin{eqnarray}
\frac{P_{\rm rad}}{P_{\rm gas}} &\simeq& 2\pi\alpha \mu^{5/2}m_p^{5/2}k_b^{-5/2}a_r(GM)^{2}\dot{M}^{-1}\left(\frac{H}{r}\right)^{6} \nonumber \\
&\approx& 1\left(\frac{\alpha}{0.1}\right)\left(\frac{H/r}{0.4}\right)^{6}\left(\frac{M}{1.4M_{\odot}}\right)^{2}\left(\frac{T_{\rm burn}}{10^{9}{\rm K}}\right)^{3/2}\left(\frac{\dot{M}}{10^{-3}M_{\odot}{\,\rm s^{-1}}}\right)^{-1}.
\label{eq:pratio}
\end{eqnarray} 

Turbulent mixing likely sets a lower limit of $\Delta r_{\rm burn} \gtrsim H$ on the radial extent of burning.  On the other hand, burning is unlikely to occur over a much larger radial region $\Delta r_{\rm burn} \gg r_{\rm burn} \sim 2H$ due to the sensitive temperature dependence of the nuclear reaction rates and the power-law behavior of $T(r)$.  The bracketed quantity in equation (\ref{eq:qdotratio}) is thus a reasonable estimate of $\dot{q}_{\rm nuc}/\dot{q}_{\rm visc}$.  The approximate equality $\dot{q}_{\rm nuc} \gtrsim \dot{q}_{\rm visc}$ may be considered a loose definition for whether a disk is in the NuDAF regime on a radial scale $r \approx r_{\rm burn}$.

As an example, consider carbon burning ($Q/A \simeq 0.38$ MeV nucleon$^{-1}$; $T_{\rm burn} \approx 1.5\times 10^{9}$ K) in a NS disk with a mass fraction $X_{12} = 0.5$, for which $\Delta\epsilon_{\rm nuc} \approx 0.2$ MeV.  Equation (\ref{eq:qdotratio}) shows that $\dot{q}_{\rm nuc} \lesssim \dot{q}_{\rm visc}$ for $\dot{M} = 10^{-3}M_{\odot}$ s$^{-1}$ and $\alpha = 0.1$, a result seemingly independent of $\dot{M}$ or the mass of the central object.  However, this result holds only if gas pressure dominates ($\dot{M} \gtrsim 10^{-3}(\alpha/0.1)M_{\odot}$ s$^{-1}$; eq.~[\ref{eq:qdotratio}]).  At lower accretion rates, radiation pressure dominates, in which case $\dot{q}_{\rm nuc}/\dot{q}_{\rm visc}$ decreases $\propto \dot{M}^{0.4}M^{-0.8}$.

\section{Electromagnetic Counterparts}
\label{sec:observe}
In this section we use our results from $\S\ref{sec:results}$ to assess possible electromagnetic (EM) counterparts of WD-NS/BH mergers.

\subsection{Gamma-Ray Burst}
\label{sec:GRB}

A previously discussed EM counterpart of WD-NS/BH mergers is a long duration Gamma-Ray Burst, powered by accretion onto the central BH \citep{Fryer+99} or NS \citep{King+07}.  Although the formation of a relativistic jet from the inner disk is certainly possible, our calculations show that the accreted mass reaching the NS surface or BH event horizon may be a factor $\sim 10-100$ smaller than that of the original WD (although see $\S\ref{sec:convection}$).  Among other things, this suggests that WD-NS mergers resulting in BH formation may be relatively rare.

Even if sufficient mass accretes to produce a powerful jet, only a small fraction of GRB jets are pointed towards the Earth.  For off-axis events, the prompt and afterglow emission are much dimmer due to relativistic debeaming.  Thus, although it would be unsurprising if WD-NS/BH mergers were accompanied by non-thermal jetted emission at some level, bright high energy emission may not be a ubiquitous feature.

\subsection{Radioactively-Powered Optical Transient}
\label{sec:SN}
\begin{figure}
\resizebox{\hsize}{!}{\includegraphics[angle=0]{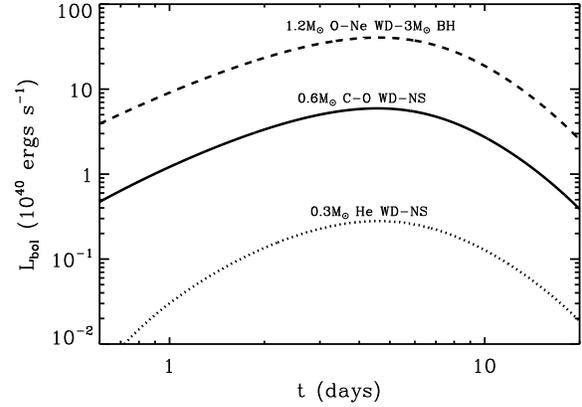}}
\caption[] {Bolometric light curve of supernova-like emission from WD-NS/BH mergers, powered by the decay of $^{56}$Ni in the wind-driven ejecta.  Three models shown correspond to the merger of [1] a 0.6$M_{\odot}$ C-O WD with a 1.4$M_{\odot}$ NS ({\it solid}; $\bar{v}_{\rm w} \simeq 2.8\times 10^{4}$ km s$^{-1}$; $M_{\rm ej} \simeq 0.5M_{\odot}$; $M_{\rm Ni} = 6\times 10^{-3}M_{\odot}$; Fig.~\ref{fig:CO}); [2] a 0.3$M_{\odot}$ He WD with a 1.2$M_{\odot}$ NS ({\it dotted}; $\bar{v}_{\rm w} \simeq 2.2\times 10^{4}$ km s$^{-1}$; $M_{\rm ej} \simeq 0.3M_{\odot}$; $M_{\rm Ni} = 3\times 10^{-4}M_{\odot}$; Fig.~\ref{fig:He}); and [3] a 1.2$M_{\odot}$ O-Ne WD with a 3$M_{\odot}$ BH ({\it dashed}; $\bar{v}_{\rm w} \simeq 4.6\times 10^{4}$ km s$^{-1}$; $M_{\rm ej} \simeq 1M_{\odot}$; $M_{\rm Ni} = 5\times 10^{-2}M_{\odot}$; Fig.~\ref{fig:ONe})
}
\label{fig:SN}
\end{figure}

A promising source of isotropic EM emission from WD-NS/BH mergers is a supernova(SN)-like transient, powered by the radioactive decay of $^{56}$Ni from the wind-driven ejecta.  We argued in $\S\ref{sec:outflowresults}$ that outflows from the disk form a singular homologous `shell' on large scales.  Depending on the binary parameters and the wind efficiency $\eta_{\rm w}$, the ejecta has a characteristic mass $M_{\rm ej} \sim 0.3-1M_{\odot}$, mean velocity $\bar{v}_{\rm ej} \approx 1-5\times 10^{4}$ km s$^{-1}$, and Ni mass $M_{\rm Ni} \approx X_{56}M_{\rm ej} \sim 10^{-4.5}-10^{-2.5}M_{\odot}$ (Table \ref{table:results}).  More massive WDs generally produce more massive ejecta, with a higher speed and larger Ni yield.  This conclusion should be taken with caution, however; thermal instability may alter the disk evolution, especially for high mass (and possibly hybrid) WDs ($\S\ref{sec:stability}$).  

%SN emission peaks when the timescale for photon diffusion through the ejecta $t_{\rm d} = B\kappa M_{\rm ej}/Rc$ equals the expansion time $t_{\rm exp} = R/\bar{v}_{\rm ej}$ \citep{Arnett82}, where $R$ is the mean radius of the ejecta, $\kappa$ is the opacity, and $B$ is a geometric factor $\simeq 0.07$ for a spherical outflow (e.g. \citealt{Padmanabhan:2000}).  This occurs on a timescale
%\be
%t_{\rm peak} \approx 5{\rm\, days}\left(\frac{v}{3\times 10^{4}{\,\rm km\,s^{-1}}}\right)^{-1/2}\left(\frac{M_{\rm ej}}{M_{\sun}}\right)^{1/2}.
%\label{eq:tpeak}
%\ee 
%where we take $\kappa = 0.1$ cm$^{2}$ g$^{-1}$ as an estimate of the opacity \citep{Pinto.Eastman:2000}.  The peak bolometric luminosity, by contrast, matches that produced instantaneously by radioactive Ni decay, viz.~
%\be L_{\rm peak} \simeq 8\times 10^{41}\epsilon_{\rm th}\left(\frac{M_{\rm Ni}}{10^{-2}M_{\odot}}\right){\,\rm ergs\,s^{-1}},
%\label{eq:Lpeak}
%\ee  
%where $\epsilon_{\rm th} \sim 0.1-1$ is the fraction of the $\gamma-$ray energy that thermalizes with the ejecta (Colgate et al.~1980), and we have assumed that $t_{\rm peak}\lesssim $ the 6-day $^{56}$Ni half-life.

Figure \ref{fig:SN} shows bolometric light curves of Ni decay-powered transients from WD-NS/BH mergers, calculated using the model of \citet{Kulkarni05} and \citet{Metzger+08} for ejecta properties corresponding to different examples of WD-NS/BH systems from $\S\ref{sec:results}$.  In all cases the light curve peaks on a timescale $\sim$ week.  By contrast, the peak luminosity $L_{\rm bol}$ varies from $\sim 10^{39}$ ergs s$^{-1}$ to $\sim 10^{41.5}$ ergs s$^{-1}$ (absolute magnitude $M_{\rm B} \simeq -9$ to $\simeq -15$), as results primarily from the large spread in $^{56}$Ni mass, which depends primarily on the mass of the disrupted WD. 

\subsubsection{Comparison to observed subluminous Type I SN}
\label{sec:subluminous}

At least in the case of C-O and O-Ne mergers, both the brightness and duration of the predicted transient are broadly consistent with the properties of recently discovered subluminous Type I SNe such as 2005E, 2008ha, and 2010X ($\S\ref{sec:intro}$).  On closer inspection, however, connecting individual events (or classes of events) to WD-NS/BH mergers results in several potential difficulties.  

Although the ejecta mass, Ni mass, and presence of C, O, Si, S, Ca, Fe from C-O/O-Ne WD mergers are all broadly consistent with the observed properties of SN 2008ha (\citealt{Valenti+09}; \citealt{Foley+09,Foley+10}), the velocity of the ejecta $\lesssim 10^{4}$ km s$^{-1}$ inferred for this and other `2002cx-like' events are much lower than the predictions of our baseline models.  One caveat is that our calculations of $\bar{v}_{\rm ej}$ assume that winds from all radii in the disk contribute to a single homologous body of ejecta, whereas in reality $\bar{v}_{\rm ej}$ could be smaller if the fast wind from the inner disk escapes along the pole without efficiently coupling its kinetic energy.  

The ejecta mass, Ni mass, velocity, and presence of He, Ca, and O of our C-O and He WD models are similarly consistent with the properties of SN 2005E.  However, the large Ca mass $\gtrsim 0.1M_{\odot}$ inferred in the ejecta of this and related events \citep{Perets+10} is much higher than we predict.  If this disagreement can be reconciled, we speculate that the mysterious location of 05E-like objects in the outskirts of their host galaxies could be explained by our model if WD-NS binaries are given a `kick' during their supernova, which removes them from the disk of the galaxy by the time of merger.    

For SN 2010X, again many of the elements produced by WD-NS/BH mergers are seen in the spectra (and some, like C and other intermediate mass elements, are not expected in alternative .Ia models; \citealt{Kasliwal+10}).  However, Ti is also observed, despite its low predicted quantity in our models.  Furthermore, none of our models produce enough $^{56}$Ni to explain SN 2002bj \citep{Poznanski+10}, an event which otherwise shares several properties with 2010X.  Again, these conclusions must be moderated due to our ignorance of the outcome of high mass ratio mergers and the uncertain effects of thermal instabilities.

As a final note, although this possibility has been neglected thus far, free nucleons from the very inner disk may also contribute a small fraction of the ejected mass (Fig.~\ref{fig:cartoon}).  These may be lost to winds heated by viscous dissipation (as discussed here) or by neutrino irradiation from the very inner disk or boundary layer (e.g.~\citealt{Metzger+08a}, \citealt{Metzger+08b}).  Once neutrons decay, the net effect is contamination of the ejecta with hydrogen.  Since even a small quantity of H may be detectable due to its strong lines, its presence in an event otherwise classified `Type I' would support the WD-NS/BH merger model, due to its unlikely presence in other WD models (its absence would, however, not rule out the model).

\subsection{Radio Transients}
\label{sec:radio}

A final source of transient emission from WD-NS/BH mergers is non-thermal synchrotron emission powered by the deceleration of the ejecta with the surrounding interstellar medium (ISM).  Shock emission peaks once the ejecta sweeps up its own mass in the ISM; this occurs on a timescale
\be
t_{\rm dec} \approx  10\left(\frac{M_{\rm ej}}{0.3M_{\odot}}\right)^{1/3}\left(\frac{\bar{v}_{\rm ej}}{10^{4}{\rm\,km\,s^{-1}}}\right)^{-1}\left(\frac{n}{\rm cm^{-3}}\right)^{-1/3}{\rm\,yrs},
\label{eq:tdec}
\ee
where $n$ is the ISM density.  Following \citet{Nakar&Piran11} (their equation [14]), we estimate that the number of radio transients detectable with a single 1.4 GHz snapshot of the whole sky down to a limiting flux $F_{\rm lim}$ is given by
\begin{eqnarray}
&&N_{1.4} \approx 10\left(\frac{M_{\rm ej}}{0.3M_{\odot}}\right)^{1.8}\left(\frac{\bar{v}_{\rm ej}}{10^{4}{\rm km\,s^{-1}}}\right)^{6.1}\times \nonumber \\
&&\left(\frac{n}{\rm cm^{-3}}\right)\left(\frac{\epsilon_{B}}{0.1}\right)^{1.3}\left(\frac{\epsilon_{e}}{0.1}\right)^{2.3}\left(\frac{F_{\rm lim}}{0.1\rm\,mJy}\right)^{-1.5}\left(\frac{\mathcal{R}}{10^{-4}\rm\,yr^{-1}}\right)
\label{eq:N}
\end{eqnarray}
where $\mathcal{R} \sim 10^{-5}-10^{-3}$ yr$^{-1}$ is the merger rate per galaxy (see $\S\ref{sec:unstable}$) and we have assumed that relativistic electrons are accelerated into a power-law energy distribution with index $p = 2.5$.  Here $\epsilon_{e}(\epsilon_{B})$ are the fraction of the energy density behind the shock imparted to relativistic electrons and magnetic fields, respectively, normalized to values characteristic of radio supernovae \citep{Chevalier98}.  Equations (\ref{eq:tdec}) and (\ref{eq:N}) show that future wide-field radio surveys at $\sim$ GHz frequencies (e.g.~\citealt{Bower+10}) will preferentially detect those events with the highest ejecta mass and velocity, as characterize higher mass WD mergers in our model.  This is due both to the sensitive dependence of $N_{1.4}$ on $M_{\rm ej}$ and $\bar{v}_{\rm ej}$ and the requirement that $t_{\rm dec}$ be sufficiently short to produce an appreciable change in the radio brightness over the characteristic lifetime of the survey.

\section{Discussion and Conclusions}
\label{sec:conclusion}

Although the effects of nuclear burning on accretion have been explored in previous work (e.g.~\citealt{Taam&Fryxell85}; \citealt{Chakrabarti+87}), these efforts focused on systems with much lower accretion rates, such as X-ray binaries.  Unphysically low values of the disk viscosity $\alpha \lesssim 10^{-10}$ are necessary in these cases to achieve sufficiently high densities for appreciable nucleosynthesis.  Here we have shown that nuclear burning has an important effect on the {\it dynamics} of accretion following the tidal disruption of a WD by a NS or stellar mass BH, even for more realistic values of $\alpha \gtrsim 0.01-0.1$.  Under conditions when the heating from nuclear burning is comparable to, or greater than that, released by viscous dissipation, we have introduced the concept of a `Nuclear-Dominated Accretion Flow' (NuDAF; see the discussion surrounding eq.~[\ref{eq:qdotratio}] for a more concrete definition of this regime).  The dependence of the disk thermodynamics on nuclear burning is particularly acute because inflow is already radiatively-inefficient and hence marginally bound \citep{Narayan&Yi94}, even without an additional heat source.

In $\S\ref{sec:observe}$ we discussed several sources of transient emission associated with WD-NS/BH mergers.  In addition to their EM signatures, WD-NS/BH binaries are considered a promising source of gravitational wave emission at frequencies $\lesssim $mHz.  Estimates suggest that a spaced-based interferometer such as LISA could resolve $\sim 1-100$ WD-NS/BH systems within our galaxy (\citealt{Nelemans+01}; \citealt{Cooray04}; \citealt{Paschalidis+09}).  Unfortunately, most of the discovered systems will require $\gtrsim 10^{4}$ years to merge.  Although the odds are not favorable, if a WD-NS/BH binary in the Milky Way or a nearby galaxy were sufficiently compact to merge on a shorter timescale, it should be relatively easy to detect and localize.  EM counterpart searches could then be triggered at the expected merger time.

The results presented here may be applied to other contexts.  One immediate possibility is the tidal disruption of a WD by an intermediate mass BH (e.g.~\citealt{Rosswog+09}; \citealt{Clausen&Eracleous11}).  For nearly circular orbits prior to merger (as we have discussed), burning is unlikely to be important if the primary mass $M$ is too large.  This is because when radiation-pressure dominates (as is likely in this case; eq.~[\ref{eq:pratio}]), the ratio of nuclear to viscous heating $\dot{q}_{\rm nuc}/\dot{q}_{\rm visc} \propto M^{-0.8}$ (eq.~[\ref{eq:qdotratio}]) is small due to the deep potential well of the BH.  On the other hand, for highly elliptical orbits with small pericenter radii, tidal forces during the disruption can themselves trigger a nuclear runaway (\citealt{Rosswog+09}); because the initial orbit has a very low binding energy, the nuclear energy released will have a more significant effect on what matter is unbound.

A perhaps more promising application of our results is to accretion following the core collapse of a massive star, as in the `collapsar' model for long duration GRBs (\citealt{MacFadyen&Woosley99}).  Notably, the outer layers of the Wolf-Rayet progenitors of long GRBs are predicted to have a He-C-O composition \citep{Heger&Woosley06}, similar to that of a `hybrid' WD (Fig.~[\ref{fig:hybrid}]).  Due to its low temperature threshold, the reaction $^{4}$He+$^{16}$O$\rightarrow$ $^{20}$Ne+$\gamma$ may have a particularly important influence on the disk, if material circularizes at a sufficiently large radius.  Note that current numerical simulations of collapsar disks generally neglect the effects of nuclear reactions, except perhaps at very high temperature $\gtrsim 4\times 10^{9}$ K (\citealt{MacFadyen&Woosley99}; \citealt{Lindner+10}; \citealt{Milosavljevic+10}).  Our results show that may not be valid, as nuclear energy generation can be important, even for temperatures as low as $\sim 10^{9}$ K.  The effects of nuclear burning may be similarly important in accretion following the merger of a helium star with a NS or BH \citep{Fryer&Woosley98}.  The recent GRB 101225A was interpreted as resulting from accretion following following NS-helium star merger \citep{Thone+11}; we note that the small Ni mass inferred from the weak SN associated with this event is consistent with our estimated yield from He WD-NS mergers.

There is much room for improvement on the model presented here in future work.  For one, we have assumed a time-steady, height-integrated disk model.  In reality, the disk may have a complex vertical structure, due in part to the interplay between MRI-driven turbulence and convection, the latter driven by the strong temperature dependence of nuclear heating.  Complex time-dependent behavior is also expected, associated with both the secular viscous evolution of the disk and, potentially, shorter timescale variability associated with thermal instability (possibly resulting in limit cycle behavior; $\S\ref{sec:stability}$).  We have assumed a simple outflow model with two uncertain parameters, $\eta_{\rm w}$ and Be$_{\rm d}^{'}$, whose values depend on the mechanism and heating source responsible for driving outflows from the disk corona.  The pressure dependence of wind cooling has important implications for the thermal stability of the disk ($\S\ref{sec:stability}$).  Our calculations also neglect the effects of convection on energy and angular momentum transport in the disk ($\S\ref{sec:convection}$), which, if important, could substantially increase the accretion timescale or even alter the radial structure of the accreting envelope.  Even our simple model cannot be directly applied to the merger of massive WDs with NSs, since nuclear burning is already important during the disk's formation itself.  Here, multi-dimensional numerical simulations that include the effects of nuclear reactions will be required to fully characterize the evolution and fate of these systems. 

Finally, the discussion of optical and radio EM counterparts in $\S\ref{sec:SN}$ and $\S\ref{sec:radio}$ was necessarily limited in scope.  In future work we will pursue more detailed calculations of the predicted optical light curve and spectra, as will be necessary to more precisely compare the predictions of our model to observations of subluminous Type I SNe ($\S\ref{sec:subluminous}$).  This work will also address the expected detection rates with current and upcoming optical and radio transient surveys, given the current constraints on the merger rates of different WD-NS/BH systems.

\section*{Acknowledgments}

I thank J.~Goodman for many helpful conservations and for encouraging my work on this topic, as well as for a thorough reading of the text.  I thank C.~Kim and D.~Lorimer for helpful information on WD-NS binaries.  I also thank D.~Giannios, E.~Quataert, R.~Narayan, A.~Piro, R.~Foley, M.~Kasliwal, and L.~Bildsten for helpful conversations and information.  I thank F.~Timmes for making his nuclear reaction codes available to the public.  BDM is supported by NASA through Einstein Postdoctoral Fellowship grant number PF9-00065 awarded by the Chandra X-ray Center, which is operated by the Smithsonian Astrophysical Observatory for NASA under contract NAS8-03060.

\bibliographystyle{mn2e}
%\bibliography{../biblio/bibliography}
\bibliography{ms}

\begin{thebibliography}{}

\bibitem[\protect\citeauthoryear{{Bailes}, {Ord}, {Knight} \& {Hotan}}{{Bailes}
  et~al.}{2003}]{Bailes+03}
{Bailes} M.,  {Ord} S.~M.,  {Knight} H.~S.,    {Hotan} A.~W.,  2003, \apjl,
  595, L49

\bibitem[\protect\citeauthoryear{{Begelman}, {Rossi} \& {Armitage}}{{Begelman}
  et~al.}{2008}]{Begelman+08}
{Begelman} M.~C.,  {Rossi} E.~M.,    {Armitage} P.~J.,  2008, \mnras, 387, 1649

\bibitem[\protect\citeauthoryear{{Bildsten}, {Shen}, {Weinberg} \&
  {Nelemans}}{{Bildsten} et~al.}{2007}]{Bildsten+07}
{Bildsten} L.,  {Shen} K.~J.,  {Weinberg} N.~N.,    {Nelemans} G.,  2007,
  \apjl, 662, L95

\bibitem[\protect\citeauthoryear{{Blandford} \& {Begelman}}{{Blandford} \&
  {Begelman}}{1999}]{Blandford&Begelman99}
{Blandford} R.~D.,  {Begelman} M.~C.,  1999, \mnras, 303, L1

\bibitem[\protect\citeauthoryear{{Bower} et~al.,}{{Bower}
  et~al.}{2010}]{Bower+10}
{Bower} G.~C.,  et~al., 2010, \apj, 725, 1792

\bibitem[\protect\citeauthoryear{{Branch}, {Baron}, {Thomas}, {Kasen}, {Li} \&
  {Filippenko}}{{Branch} et~al.}{2004}]{Branch+04}
{Branch} D.,  {Baron} E.,  {Thomas} R.~C.,  {Kasen} D.,  {Li} W.,
  {Filippenko} A.~V.,  2004, \pasp, 116, 903

\bibitem[\protect\citeauthoryear{{Cannizzo}}{{Cannizzo}}{1993}]{Cannizzo93}
{Cannizzo} J.~K.,  1993, \apj, 419, 318

\bibitem[\protect\citeauthoryear{{Carballido}, {Stone} \&
  {Pringle}}{{Carballido} et~al.}{2005}]{Carballido+05}
{Carballido} A.,  {Stone} J.~M.,    {Pringle} J.~E.,  2005, \mnras, 358, 1055

\bibitem[\protect\citeauthoryear{{Chakrabarti}, {Jin} \&
  {Arnett}}{{Chakrabarti} et~al.}{1987}]{Chakrabarti+87}
{Chakrabarti} S.~K.,  {Jin} L.,    {Arnett} W.~D.,  1987, \apj, 313, 674

\bibitem[\protect\citeauthoryear{{Chen} \& {Beloborodov}}{{Chen} \&
  {Beloborodov}}{2007}]{Chen&Beloborodov07}
{Chen} W.-X.,  {Beloborodov} A.~M.,  2007, \apj, 657, 383

\bibitem[\protect\citeauthoryear{{Chevalier}}{{Chevalier}}{1998}]{Chevalier98}
{Chevalier} R.~A.,  1998, \apj, 499, 810

\bibitem[\protect\citeauthoryear{{Clausen} \& {Eracleous}}{{Clausen} \&
  {Eracleous}}{2011}]{Clausen&Eracleous11}
{Clausen} D.,  {Eracleous} M.,  2011, \apj, 726, 34

\bibitem[\protect\citeauthoryear{{Cooray}}{{Cooray}}{2004}]{Cooray04}
{Cooray} A.,  2004, \mnras, 354, 25

\bibitem[\protect\citeauthoryear{{Davies}, {Ritter} \& {King}}{{Davies}
  et~al.}{2002}]{Davies+02}
{Davies} M.~B.,  {Ritter} H.,    {King} A.,  2002, \mnras, 335, 369

\bibitem[\protect\citeauthoryear{{Davis}, {Stone} \& {Pessah}}{{Davis}
  et~al.}{2010}]{Davis+10}
{Davis} S.~W.,  {Stone} J.~M.,    {Pessah} M.~E.,  2010, \apj, 713, 52

\bibitem[\protect\citeauthoryear{{D'Souza}, {Motl}, {Tohline} \&
  {Frank}}{{D'Souza} et~al.}{2006}]{DSouza+06}
{D'Souza} M.~C.~R.,  {Motl} P.~M.,  {Tohline} J.~E.,    {Frank} J.,  2006,
  \apj, 643, 381

\bibitem[\protect\citeauthoryear{{Edwards} \& {Bailes}}{{Edwards} \&
  {Bailes}}{2001}]{Edwards&Bailes01}
{Edwards} R.~T.,  {Bailes} M.,  2001, \apjl, 547, L37

\bibitem[\protect\citeauthoryear{{Eggleton}}{{Eggleton}}{1983}]{Eggleton83}
{Eggleton} P.~P.,  1983, \apj, 268, 368

\bibitem[\protect\citeauthoryear{{Foley}, {Brown}, {Rest}, {Challis},
  {Kirshner} \& {Wood-Vasey}}{{Foley} et~al.}{2010}]{Foley+10}
{Foley} R.~J.,  {Brown} P.~J.,  {Rest} A.,  {Challis} P.~J.,  {Kirshner} R.~P.,
     {Wood-Vasey} W.~M.,  2010, \apjl, 708, L61

\bibitem[\protect\citeauthoryear{{Foley}, {Chornock}, {Filippenko},
  {Ganeshalingam}, {Kirshner}, {Li}, {Cenko}, {Challis}, {Friedman}, {Modjaz},
  {Silverman} \& {Wood-Vasey}}{{Foley} et~al.}{2009}]{Foley+09}
{Foley} R.~J.,  {Chornock} R.,  {Filippenko} A.~V.,  {Ganeshalingam} M.,
  {Kirshner} R.~P.,  {Li} W.,  {Cenko} S.~B.,  {Challis} P.~J.,  {Friedman}
  A.~S.,  {Modjaz} M.,  {Silverman} J.~M.,    {Wood-Vasey} W.~M.,  2009, AJ,
  138, 376

\bibitem[\protect\citeauthoryear{{Freire} \& {Wex}}{{Freire} \&
  {Wex}}{2010}]{Freire&Wex10}
{Freire} P.,  {Wex} N.,  2010, ArXiv e-prints

\bibitem[\protect\citeauthoryear{{Fryer} \& {Woosley}}{{Fryer} \&
  {Woosley}}{1998}]{Fryer&Woosley98}
{Fryer} C.~L.,  {Woosley} S.~E.,  1998, \apjl, 502, L9+

\bibitem[\protect\citeauthoryear{{Fryer}, {Woosley}, {Herant} \&
  {Davies}}{{Fryer} et~al.}{1999}]{Fryer+99}
{Fryer} C.~L.,  {Woosley} S.~E.,  {Herant} M.,    {Davies} M.~B.,  1999, \apj,
  520, 650

\bibitem[\protect\citeauthoryear{{Guerrero}, {Garc{\'{\i}}a-Berro} \&
  {Isern}}{{Guerrero} et~al.}{2004}]{Guerrero+04}
{Guerrero} J.,  {Garc{\'{\i}}a-Berro} E.,    {Isern} J.,  2004, \aap, 413, 257

\bibitem[\protect\citeauthoryear{{Guti{\'e}rrez}, {Canal} \&
  {Garc{\'{\i}}a-Berro}}{{Guti{\'e}rrez} et~al.}{2005}]{Gutierrez+05}
{Guti{\'e}rrez} J.,  {Canal} R.,    {Garc{\'{\i}}a-Berro} E.,  2005, \aap, 435,
  231

\bibitem[\protect\citeauthoryear{{Han}, {Tout} \& {Eggleton}}{{Han}
  et~al.}{2000}]{Han+00}
{Han} Z.,  {Tout} C.~A.,    {Eggleton} P.~P.,  2000, \mnras, 319, 215

\bibitem[\protect\citeauthoryear{{Hawley} \& {Balbus}}{{Hawley} \&
  {Balbus}}{2002}]{Hawley&Balbus02}
{Hawley} J.~F.,  {Balbus} S.~A.,  2002, \apj, 573, 738

\bibitem[\protect\citeauthoryear{{Hawley}, {Balbus} \& {Stone}}{{Hawley}
  et~al.}{2001}]{Hawley+01}
{Hawley} J.~F.,  {Balbus} S.~A.,    {Stone} J.~M.,  2001, \apjl, 554, L49

\bibitem[\protect\citeauthoryear{{Iben} Jr. \& {Tutukov}}{{Iben} \&
  {Tutukov}}{1991}]{Iben&Tutukov91}
{Iben} Jr. I.,  {Tutukov} A.~V.,  1991, \apj, 370, 615

\bibitem[\protect\citeauthoryear{{Igumenshchev}, {Abramowicz} \&
  {Narayan}}{{Igumenshchev} et~al.}{2000}]{Igumenshchev+00}
{Igumenshchev} I.~V.,  {Abramowicz} M.~A.,    {Narayan} R.,  2000, \apjl, 537,
  L27

\bibitem[\protect\citeauthoryear{{Jha}, {Branch}, {Chornock}, {Foley}, {Li},
  {Swift}, {Casebeer} \& {Filippenko}}{{Jha} et~al.}{2006}]{Jha+06}
{Jha} S.,  {Branch} D.,  {Chornock} R.,  {Foley} R.~J.,  {Li} W.,  {Swift}
  B.~J.,  {Casebeer} D.,    {Filippenko} A.~V.,  2006, \aj, 132, 189

\bibitem[\protect\citeauthoryear{{Kaiser} et~al.,}{{Kaiser}
  et~al.}{2002}]{Kaiser+02}
{Kaiser} N.,  et~al., 2002, in {J.~A.~Tyson \& S.~Wolff} ed., Society of
  Photo-Optical Instrumentation Engineers (SPIE) Conference Series Vol.~4836 of
  Society of Photo-Optical Instrumentation Engineers (SPIE) Conference Series,
  {Pan-STARRS: A Large Synoptic Survey Telescope Array}.
pp 154--164

\bibitem[\protect\citeauthoryear{{Kasliwal} et~al.,}{{Kasliwal}
  et~al.}{2010}]{Kasliwal+10}
{Kasliwal} M.~M.,  et~al., 2010, \apjl, 723, L98

\bibitem[\protect\citeauthoryear{{Kaspi}, {Lyne}, {Manchester}, {Crawford},
  {Camilo}, {Bell}, {D'Amico}, {Stairs}, {McKay}, {Morris} \&
  {Possenti}}{{Kaspi} et~al.}{2000}]{Kaspi+00}
{Kaspi} V.~M.,  {Lyne} A.~G.,  {Manchester} R.~N.,  {Crawford} F.,  {Camilo}
  F.,  {Bell} J.~F.,  {D'Amico} N.,  {Stairs} I.~H.,  {McKay} N.~P.~F.,
  {Morris} D.~J.,    {Possenti} A.,  2000, \apj, 543, 321

\bibitem[\protect\citeauthoryear{{Kim}, {Kalogera}, {Lorimer} \& {White}}{{Kim}
  et~al.}{2004}]{Kim+04}
{Kim} C.,  {Kalogera} V.,  {Lorimer} D.~R.,    {White} T.,  2004, \apj, 616,
  1109

\bibitem[\protect\citeauthoryear{{King}, {Olsson} \& {Davies}}{{King}
  et~al.}{2007}]{King+07}
{King} A.,  {Olsson} E.,    {Davies} M.~B.,  2007, \mnras, 374, L34

\bibitem[\protect\citeauthoryear{{King}, {Pringle} \& {Livio}}{{King}
  et~al.}{2007}]{King+07b}
{King} A.~R.,  {Pringle} J.~E.,    {Livio} M.,  2007, \mnras, 376, 1740

\bibitem[\protect\citeauthoryear{{Kohri}, {Narayan} \& {Piran}}{{Kohri}
  et~al.}{2005}]{Kohri+05}
{Kohri} K.,  {Narayan} R.,    {Piran} T.,  2005, \apj, 629, 341

\bibitem[\protect\citeauthoryear{{Kulkarni}}{{Kulkarni}}{2005}]{Kulkarni05}
{Kulkarni} S.~R.,  2005, ArXiv Astrophysics e-prints

\bibitem[\protect\citeauthoryear{{Lamers} \& {Cassinelli}}{{Lamers} \&
  {Cassinelli}}{1999}]{Lamers&Cassinelli99}
{Lamers} H.~J.~G.~L.~M.,  {Cassinelli} J.~P.,  1999, {Introduction to Stellar
  Winds}

\bibitem[\protect\citeauthoryear{{Laughlin} \& {Bodenheimer}}{{Laughlin} \&
  {Bodenheimer}}{1994}]{Laughlin&Bodenheimer94}
{Laughlin} G.,  {Bodenheimer} P.,  1994, \apj, 436, 335

\bibitem[\protect\citeauthoryear{{Law} et~al.,}{{Law}  et~al.}{2009}]{Law+09}
{Law} N.~M.,  et~al., 2009, PASP, 121, 1395

\bibitem[\protect\citeauthoryear{{Li}, {Filippenko}, {Chornock}, {Berger},
  {Berlind}, {Calkins}, {Challis}, {Fassnacht}, {Jha}, {Kirshner}, {Matheson},
  {Sargent}, {Simcoe}, {Smith} \& {Squires}}{{Li} et~al.}{2003}]{Li+03}
{Li} W.,  {Filippenko} A.~V.,  {Chornock} R.,  {Berger} E.,  {Berlind} P.,
  {Calkins} M.~L.,  {Challis} P.,  {Fassnacht} C.,  {Jha} S.,  {Kirshner}
  R.~P.,  {Matheson} T.,  {Sargent} W.~L.~W.,  {Simcoe} R.~A.,  {Smith} G.~H.,
    {Squires} G.,  2003, \pasp, 115, 453

\bibitem[\protect\citeauthoryear{{Lindner}, {Milosavljevi{\'c}}, {Couch} \&
  {Kumar}}{{Lindner} et~al.}{2010}]{Lindner+10}
{Lindner} C.~C.,  {Milosavljevi{\'c}} M.,  {Couch} S.~M.,    {Kumar} P.,  2010,
  \apj, 713, 800

\bibitem[\protect\citeauthoryear{{Livne} \& {Arnett}}{{Livne} \&
  {Arnett}}{1995}]{Livne&Arnett95}
{Livne} E.,  {Arnett} D.,  1995, \apj, 452, 62

\bibitem[\protect\citeauthoryear{{Lorimer}}{{Lorimer}}{2005}]{Lorimer05}
{Lorimer} D.~R.,  2005, Living Reviews in Relativity, 8, 7

\bibitem[\protect\citeauthoryear{{Lundgren}, {Zepka} \& {Cordes}}{{Lundgren}
  et~al.}{1995}]{Lundgren+95}
{Lundgren} S.~C.,  {Zepka} A.~F.,    {Cordes} J.~M.,  1995, \apj, 453, 419

\bibitem[\protect\citeauthoryear{{MacFadyen} \& {Woosley}}{{MacFadyen} \&
  {Woosley}}{1999}]{MacFadyen&Woosley99}
{MacFadyen} A.~I.,  {Woosley} S.~E.,  1999, \apj, 524, 262

\bibitem[\protect\citeauthoryear{{Metzger}, {Piro} \& {Quataert}}{{Metzger}
  et~al.}{2008a}]{Metzger+08}
{Metzger} B.~D.,  {Piro} A.~L.,    {Quataert} E.,  2008a, \mnras, 390, 781

\bibitem[\protect\citeauthoryear{{Metzger}, {Piro} \& {Quataert}}{{Metzger}
  et~al.}{2008b}]{Metzger+08b}
{Metzger} B.~D.,  {Piro} A.~L.,    {Quataert} E.,  2008b, \mnras, 390, 781

\bibitem[\protect\citeauthoryear{{Metzger}, {Thompson} \& {Quataert}}{{Metzger}
  et~al.}{2008}]{Metzger+08a}
{Metzger} B.~D.,  {Thompson} T.~A.,    {Quataert} E.,  2008, \apj, 676, 1130

\bibitem[\protect\citeauthoryear{{Milosavljevic}, {Lindner}, {Shen} \&
  {Kumar}}{{Milosavljevic} et~al.}{2010}]{Milosavljevic+10}
{Milosavljevic} M.,  {Lindner} C.~C.,  {Shen} R.,    {Kumar} P.,  2010, ArXiv
  e-prints

\bibitem[\protect\citeauthoryear{{Moriya}, {Tominaga}, {Tanaka}, {Nomoto},
  {Sauer}, {Mazzali}, {Maeda} \& {Suzuki}}{{Moriya} et~al.}{2010}]{Moriya+10}
{Moriya} T.,  {Tominaga} N.,  {Tanaka} M.,  {Nomoto} K.,  {Sauer} D.~N.,
  {Mazzali} P.~A.,  {Maeda} K.,    {Suzuki} T.,  2010, \apj, 719, 1445

\bibitem[\protect\citeauthoryear{{Nakar} \& {Piran}}{{Nakar} \&
  {Piran}}{2011}]{Nakar&Piran11}
{Nakar} E.,  {Piran} T.,  2011, ArXiv e-prints

\bibitem[\protect\citeauthoryear{{Narayan}, {Igumenshchev} \&
  {Abramowicz}}{{Narayan} et~al.}{2000}]{Narayan+00}
{Narayan} R.,  {Igumenshchev} I.~V.,    {Abramowicz} M.~A.,  2000, \apj, 539,
  798

\bibitem[\protect\citeauthoryear{{Narayan}, {Mahadevan} \&
  {Quataert}}{{Narayan} et~al.}{1998}]{Narayan+98}
{Narayan} R.,  {Mahadevan} R.,    {Quataert} E.,  1998, in {M.~A.~Abramowicz,
  G.~Bjornsson, \& J.~E.~Pringle} ed., Theory of Black Hole Accretion Disks
  {Advection-dominated accretion around black holess}.
pp 148--+

\bibitem[\protect\citeauthoryear{{Narayan}, {Piran} \& {Kumar}}{{Narayan}
  et~al.}{2001}]{Narayan+01}
{Narayan} R.,  {Piran} T.,    {Kumar} P.,  2001, \apj, 557, 949

\bibitem[\protect\citeauthoryear{{Narayan} \& {Yi}}{{Narayan} \&
  {Yi}}{1994}]{Narayan&Yi94}
{Narayan} R.,  {Yi} I.,  1994, \apjl, 428, L13

\bibitem[\protect\citeauthoryear{{Narayan} \& {Yi}}{{Narayan} \&
  {Yi}}{1995}]{Narayan&Yi95}
{Narayan} R.,  {Yi} I.,  1995, \apj, 444, 231

\bibitem[\protect\citeauthoryear{{Nauenberg}}{{Nauenberg}}{1972}]{Nauenberg72}
{Nauenberg} M.,  1972, \apj, 175, 417

\bibitem[\protect\citeauthoryear{{Nelemans}, {Yungelson} \& {Portegies
  Zwart}}{{Nelemans} et~al.}{2001}]{Nelemans+01}
{Nelemans} G.,  {Yungelson} L.~R.,    {Portegies Zwart} S.~F.,  2001, \aap,
  375, 890

\bibitem[\protect\citeauthoryear{{Nomoto}, {Thielemann} \& {Yokoi}}{{Nomoto}
  et~al.}{1984}]{Nomoto+84}
{Nomoto} K.,  {Thielemann} F.-K.,    {Yokoi} K.,  1984, \apj, 286, 644

\bibitem[\protect\citeauthoryear{{Ohsuga}, {Mori}, {Nakamoto} \&
  {Mineshige}}{{Ohsuga} et~al.}{2005}]{Ohsuga+05}
{Ohsuga} K.,  {Mori} M.,  {Nakamoto} T.,    {Mineshige} S.,  2005, \apj, 628,
  368

\bibitem[\protect\citeauthoryear{{O'Shaughnessy} \& {Kim}}{{O'Shaughnessy} \&
  {Kim}}{2010}]{OShaughnessy&Kim10}
{O'Shaughnessy} R.,  {Kim} C.,  2010, \apj, 715, 230

\bibitem[\protect\citeauthoryear{{Paschalidis}, {MacLeod}, {Baumgarte} \&
  {Shapiro}}{{Paschalidis} et~al.}{2009}]{Paschalidis+09}
{Paschalidis} V.,  {MacLeod} M.,  {Baumgarte} T.~W.,    {Shapiro} S.~L.,  2009,
  \prd, 80, 024006

\bibitem[\protect\citeauthoryear{{Perets}, {Badenes}, {Arcavi}, {Simon} \&
  {Gal-yam}}{{Perets} et~al.}{2011}]{Perets+11}
{Perets} H.~B.,  {Badenes} C.,  {Arcavi} I.,  {Simon} J.~D.,    {Gal-yam} A.,
  2011, \apj, 730, 89

\bibitem[\protect\citeauthoryear{{Perets} et~al.,}{{Perets}
  et~al.}{2010}]{Perets+10}
{Perets} H.~B.,  et~al., 2010, \nat, 465, 322

\bibitem[\protect\citeauthoryear{{Phillips} et~al.,}{{Phillips}
  et~al.}{2007}]{Phillips+07}
{Phillips} M.~M.,  et~al., 2007, \pasp, 119, 360

\bibitem[\protect\citeauthoryear{{Piran}}{{Piran}}{1978}]{Piran78}
{Piran} T.,  1978, \apj, 221, 652

\bibitem[\protect\citeauthoryear{{Portegies Zwart} \& {Yungelson}}{{Portegies
  Zwart} \& {Yungelson}}{1999}]{PortegiesZwart&Yungelson99}
{Portegies Zwart} S.~F.,  {Yungelson} L.~R.,  1999, \mnras, 309, 26

\bibitem[\protect\citeauthoryear{{Poznanski}, {Chornock}, {Nugent}, {Bloom},
  {Filippenko}, {Ganeshalingam}, {Leonard}, {Li} \& {Thomas}}{{Poznanski}
  et~al.}{2010}]{Poznanski+10}
{Poznanski} D.,  {Chornock} R.,  {Nugent} P.~E.,  {Bloom} J.~S.,  {Filippenko}
  A.~V.,  {Ganeshalingam} M.,  {Leonard} D.~C.,  {Li} W.,    {Thomas} R.~C.,
  2010, Science, 327, 58

\bibitem[\protect\citeauthoryear{{Quataert} \& {Gruzinov}}{{Quataert} \&
  {Gruzinov}}{2000}]{Quataert&Gruzinov00}
{Quataert} E.,  {Gruzinov} A.,  2000, \apj, 539, 809

\bibitem[\protect\citeauthoryear{{Rau} et~al.,}{{Rau}  et~al.}{2009}]{Rau+09}
{Rau} A.,  et~al., 2009, \pasp, 121, 1334

\bibitem[\protect\citeauthoryear{{Rosswog}, {Ramirez-Ruiz} \& {Hix}}{{Rosswog}
  et~al.}{2009}]{Rosswog+09}
{Rosswog} S.,  {Ramirez-Ruiz} E.,    {Hix} W.~R.,  2009, \apj, 695, 404

\bibitem[\protect\citeauthoryear{{Ryu} \& {Goodman}}{{Ryu} \&
  {Goodman}}{1992}]{Ryu&Goodman92}
{Ryu} D.,  {Goodman} J.,  1992, \apj, 388, 438

\bibitem[\protect\citeauthoryear{{Salaris}, {Dominguez}, {Garcia-Berro},
  {Hernanz}, {Isern} \& {Mochkovitch}}{{Salaris} et~al.}{1997}]{Salaris+97}
{Salaris} M.,  {Dominguez} I.,  {Garcia-Berro} E.,  {Hernanz} M.,  {Isern} J.,
    {Mochkovitch} R.,  1997, \apj, 486, 413

\bibitem[\protect\citeauthoryear{{Shakura} \& {Sunyaev}}{{Shakura} \&
  {Sunyaev}}{1973}]{Shakura&Sunyaev73}
{Shakura} N.~I.,  {Sunyaev} R.~A.,  1973, \aap, 24, 337

\bibitem[\protect\citeauthoryear{{Shen}, {Kasen}, {Weinberg}, {Bildsten} \&
  {Scannapieco}}{{Shen} et~al.}{2010}]{Shen+10}
{Shen} K.~J.,  {Kasen} D.,  {Weinberg} N.~N.,  {Bildsten} L.,    {Scannapieco}
  E.,  2010, ArXiv e-prints

\bibitem[\protect\citeauthoryear{{Sigurdsson} \& {Rees}}{{Sigurdsson} \&
  {Rees}}{1997}]{Sigurdsson&Rees97}
{Sigurdsson} S.,  {Rees} M.~J.,  1997, \mnras, 284, 318

\bibitem[\protect\citeauthoryear{{Taam} \& {Fryxell}}{{Taam} \&
  {Fryxell}}{1985}]{Taam&Fryxell85}
{Taam} R.~E.,  {Fryxell} B.~A.,  1985, \apj, 294, 303

\bibitem[\protect\citeauthoryear{{Tauris} \& {Sennels}}{{Tauris} \&
  {Sennels}}{2000}]{Tauris&Sennels00}
{Tauris} T.~M.,  {Sennels} T.,  2000, \aap, 355, 236

\bibitem[\protect\citeauthoryear{{Thompson}}{{Thompson}}{2010}]{Thompson10}
{Thompson} T.~A.,  2010, ArXiv e-prints

\bibitem[\protect\citeauthoryear{{Th{\"o}ne} et~al.,}{{Th{\"o}ne}
  et~al.}{2011}]{Thone+11}
{Th{\"o}ne} C.~C.,  et~al., 2011, ArXiv e-prints

\bibitem[\protect\citeauthoryear{{Thorne} \& {Zytkow}}{{Thorne} \&
  {Zytkow}}{1975}]{Thorne&Zytkow75}
{Thorne} K.~S.,  {Zytkow} A.~N.,  1975, \apjl, 199, L19

\bibitem[\protect\citeauthoryear{{Timmes}}{{Timmes}}{1999}]{Timmes99}
{Timmes} F.~X.,  1999, \apjs, 124, 241

\bibitem[\protect\citeauthoryear{{Valenti}, {Pastorello}, {Cappellaro},
  {Benetti}, {Mazzali}, {Manteca}, {Taubenberger}, {Elias-Rosa}, {Ferrando},
  {Harutyunyan}, {Hentunen}, {Nissinen}, {Pian}, {Turatto}, {Zampieri} \&
  {Smartt}}{{Valenti} et~al.}{2009}]{Valenti+09}
{Valenti} S.,  {Pastorello} A.,  {Cappellaro} E.,  {Benetti} S.,  {Mazzali}
  P.~A.,  {Manteca} J.,  {Taubenberger} S.,  {Elias-Rosa} N.,  {Ferrando} R.,
  {Harutyunyan} A.,  {Hentunen} V.~P.,  {Nissinen} M.,  {Pian} E.,  {Turatto}
  M.,  {Zampieri} L.,    {Smartt} S.~J.,  2009, \nat, 459, 674

\bibitem[\protect\citeauthoryear{{van den Heuvel} \& {Bonsdema}}{{van den
  Heuvel} \& {Bonsdema}}{1984}]{vandenHeuvel&Bonsdema84}
{van den Heuvel} E.~P.~J.,  {Bonsdema} P.~T.~J.,  1984, \aap, 139, L16

\bibitem[\protect\citeauthoryear{{Verbunt} \& {Rappaport}}{{Verbunt} \&
  {Rappaport}}{1988}]{Verbunt&Rappaport88}
{Verbunt} F.,  {Rappaport} S.,  1988, \apj, 332, 193

\bibitem[\protect\citeauthoryear{{Verbunt} \& {van den Heuvel}}{{Verbunt} \&
  {van den Heuvel}}{1995}]{Verbunt&vandenHeuvel95}
{Verbunt} F.,  {van den Heuvel} E.~P.~J.,  1995, in {W.~H.~G.~Lewin, J.~van
  Paradijs, \& E.~P.~J.~van den Heuvel} ed., X-ray Binaries {Formation and
  evolution of neutron stars and black holes in binaries.}.
pp 457--494

\bibitem[\protect\citeauthoryear{{Waldman}, {Sauer}, {Livne}, {Perets},
  {Glasner}, {Mazzali}, {Truran} \& {Gal-Yam}}{{Waldman}
  et~al.}{2010}]{Waldman+10}
{Waldman} R.,  {Sauer} D.,  {Livne} E.,  {Perets} H.,  {Glasner} A.,  {Mazzali}
  P.,  {Truran} J.~W.,    {Gal-Yam} A.,  2010, ArXiv e-prints

\bibitem[\protect\citeauthoryear{{Woosley} \& {Heger}}{{Woosley} \&
  {Heger}}{2006}]{Heger&Woosley06}
{Woosley} S.~E.,  {Heger} A.,  2006, \apj, 637, 914

\bibitem[\protect\citeauthoryear{{Woosley} \& {Kasen}}{{Woosley} \&
  {Kasen}}{2010}]{Woosley&Kasen10}
{Woosley} S.~E.,  {Kasen} D.,  2010, ArXiv e-prints

\bibitem[\protect\citeauthoryear{{Woosley}, {Taam} \& {Weaver}}{{Woosley}
  et~al.}{1986}]{Woosley&Weaver86}
{Woosley} S.~E.,  {Taam} R.~E.,    {Weaver} T.~A.,  1986, \apj, 301, 601

\bibitem[\protect\citeauthoryear{{Yungelson}, {Nelemans} \& {van den
  Heuvel}}{{Yungelson} et~al.}{2002}]{Yungelson+02}
{Yungelson} L.~R.,  {Nelemans} G.,    {van den Heuvel} E.~P.~J.,  2002, \aap,
  388, 546

\end{thebibliography}

%\begin{thebibliography}{}

%\end{thebibliography}

\end{document}